\RequirePackage{fix-cm}
\documentclass[smallextended]{svjour3}       
\smartqed  
\usepackage{lineno,hyperref}
\usepackage{amsmath,mathtools}
\usepackage{multicol}
\usepackage{multirow}
\usepackage{amssymb}
\modulolinenumbers[5]
\usepackage{graphicx}

\begin{document}

\title{Influence Maximization in Social Networks: A Survey of Behaviour-Aware Methods}
\subtitle{}


\author{Ahmad Zareie       \and
        Rizos Sakellariou 
}


\institute{Ahmad Zareie \at
        Department of Computer Science, The University of Manchester, Manchester M13 9PL, U.K. \email{ahmad.zareie@postgrad.manchester.ac.uk}           
        \and
        Rizos Sakellariou \at
        Department of Computer Science, The University of Manchester, Manchester M13 9PL, U.K. \email{rizos@manchester.ac.uk}
}

\date{}

\maketitle

\begin{abstract}
Social networks have become an increasingly common abstraction to capture the interactions of individual users in a number of everyday activities and applications. As a result, the analysis of such networks has attracted lots of attention in the literature. Among the topics of interest, a key problem relates to identifying so-called influential users for a number of applications, which need to spread messages. Several approaches have been proposed to estimate users' influence and identify sets of influential users in social networks. A common basis of these approaches is to consider links between users, that is, structural or topological properties of the network. To a lesser extent, some approaches take into account users' behaviours or attitudes. Although a number of surveys have reviewed approaches based on structural properties of social networks, there has been no comprehensive review of approaches that take into account users' behaviour. This paper attempts to cover this gap by reviewing and proposing a taxonomy of such behaviour-aware methods to identify influential users in social networks.
\end{abstract}



\section{Introduction}
\label{Sec.Intro}
The proliferation of network technologies, including social media, has changed people's daily activities and patterns of interaction. Various forms of social networking applications
are used for different purposes: exchanging messages, broadcasting news, sharing opinions on topics of common interest, publicity, and so on. User interactions form social networks that may play an effective role in shaping an opinion, fast and widespread propagation of specific messages or news and may help establish opinions quickly. 
It has long been accepted that not all users play the same role or carry the same gravitas in social networks. Some users may be more active or influential or vital due to their behaviour or friends in the network. So-called influential users play an important role in social networks and can be crucial in helping spread messages quickly and widely. 
The \textit{Influence Maximization} (IM) problem is defined as the problem of identifying a set of users in a social network, who can influence broadly and effectively other users. This is known to be a complex problem, particularly as it requires some criterion to measure influence.

The IM problem was originally studied as an algorithmic problem by Domingos and Richardson~\cite{tarikh1,tarikh2} while Kempe, Kleiberg and Tardos~\cite{tarikh3} were the first to formulate the problem as a discrete optimization problem. 
Although different studies have been dedicated to solving the IM problem, investigating the aspects of the problem that help identify influential users~\cite{IM1,IM2} and/or predict the influence of users~\cite{Prediction1,Prediction2} is still an important research challenge due to the impact that messages propagated through social networks often have on today's society.

The most commonly used model to solve the problem is to represent a social network as a graph whose nodes represent users and edges indicate the relationships between users. Then, different criteria may be specified to measure the influence of users.
In some research, user's influence is determined by the network structure, which means that influential users are identified on the basis of topological properties of the graph. In other research, apart from network structure, users' characteristics of behaviour, such as preferences or trustworthiness are taken into account. We term the former {\em behaviour-agnostic} as opposed to {\em behaviour-aware} for the latter. In behaviour-agnostic approaches, the problem is that there is essentially a structural graph-based approach to identify influential users; differences in their individual behaviour are disregarded.  Taking users' behaviour into account can match everyday realities; this is the strength (and the additional complexity) of behaviour-aware approaches.


Several surveys have reviewed various methods to address the IM problem and discussed different challenges in identifying a set of influential users.
A set of surveys~\cite{Surv6,Surv7,Surv8,Surv15} focuses on specific elements of the problem, such as diffusion models or the simulating spreading process, but without providing a comprehensive review of various methods for IM. Another set of surveys~\cite{Surv1,Surv3,Surv4} focuses on measures applied to rank each node in terms of its influence. These measures, often derived from the network's structure, are known as centrality measures. The ranking, obtained by centrality measures, can then inform the choice of most influential users.   
Other surveys focus on the methods aiming to select a set of influential users considering the network as a whole; for example, influential nodes may be chosen so that all parts of a network are covered. In~\cite{Surv2}, the solutions of the IM problem are reviewed from an algorithmic perspective, which relies on diffusion models that simulate the spreading process. Along the same line but using a somewhat different classification is the work in~\cite{Surv5}. Various classifications of methods have been proposed in other surveys~\cite{Surv9,Surv10,Surv11,Surv12,surv13}. In~\cite{Surv9}, different structural methods are systematically evaluated and compared using a benchmark platform.
In~\cite{Surv10}, structural users' influence detection methods are categorized and 
their advantages and disadvantages are discussed. 
Another review and classification is given in~\cite{Surv11}, where some behaviour-aware methods related to reputation and trust between users are also discussed. The methods proposed to identify influential users in evolving networks are classified and reviewed in~\cite{Surv12,Surv14}. In~\cite{surv13}, some variants of the IM Problem are reviewed and the hardness of the problem under both traditional and parameterized complexity is described. 

The main characteristic of all these surveys is that they focus on behaviour-agnostic methods. Some behaviour-aware methods are briefly discussed in~\cite{Surv1,Surv2,Surv3,Surv5,Surv11} but without any in-depth classification or specific focus on behaviour. It is generally true that behaviour-agnostic methods have a longer history in tackling the IM problem; this may explain why they dominate in current work. However, the starting point of behaviour-aware methods is a more realistic formulation of the IM problem. As such, behaviour-aware methods have the potential to deliver solutions of higher relevance to real-world situations. Yet, behaviour-aware methods have not been considered in any previous survey as a class on their own, that is, as a separate family of solutions deserving its own fine-grained classification. This paper addresses this gap.

In view of the above, the contributions of this paper can be described as follows:
\begin{itemize}
    \item A proposal to categorize existing methods to solve the IM problem into behaviour-agnostic and behaviour-aware, which is motivated by the realization that behavioural characteristics of users play a key role in IM. 
    \item A taxonomy and a detailed review of behaviour-aware methods for the IM problem, which discusses the behavioural characteristics that have been taken into account to solve the IM problem, how these characteristics are modelled and how the properties of the problem are affected by taking into account these characteristics. 
    \item A discussion of challenges for the IM problem from a behaviour-aware perspective.
\end{itemize}

The rest of the paper is organized as follows. Some basic concepts of social networks and the influence maximization problem are introduced in Section~\ref{sec_primin}. An overview of our taxonomy is also given in this section. Section~\ref{sec_beha_awr} covers a comprehensive overview of behaviour-aware methods for IM. Challenges and future research directions are finally presented in Section~\ref{sec_future}.


\section{Preliminaries}
\label{sec_primin}

\subsection{Problem Description}
\label{sec_PDescr}

\begin{figure}[hbt]
\begin{center}
\includegraphics[scale=0.60]{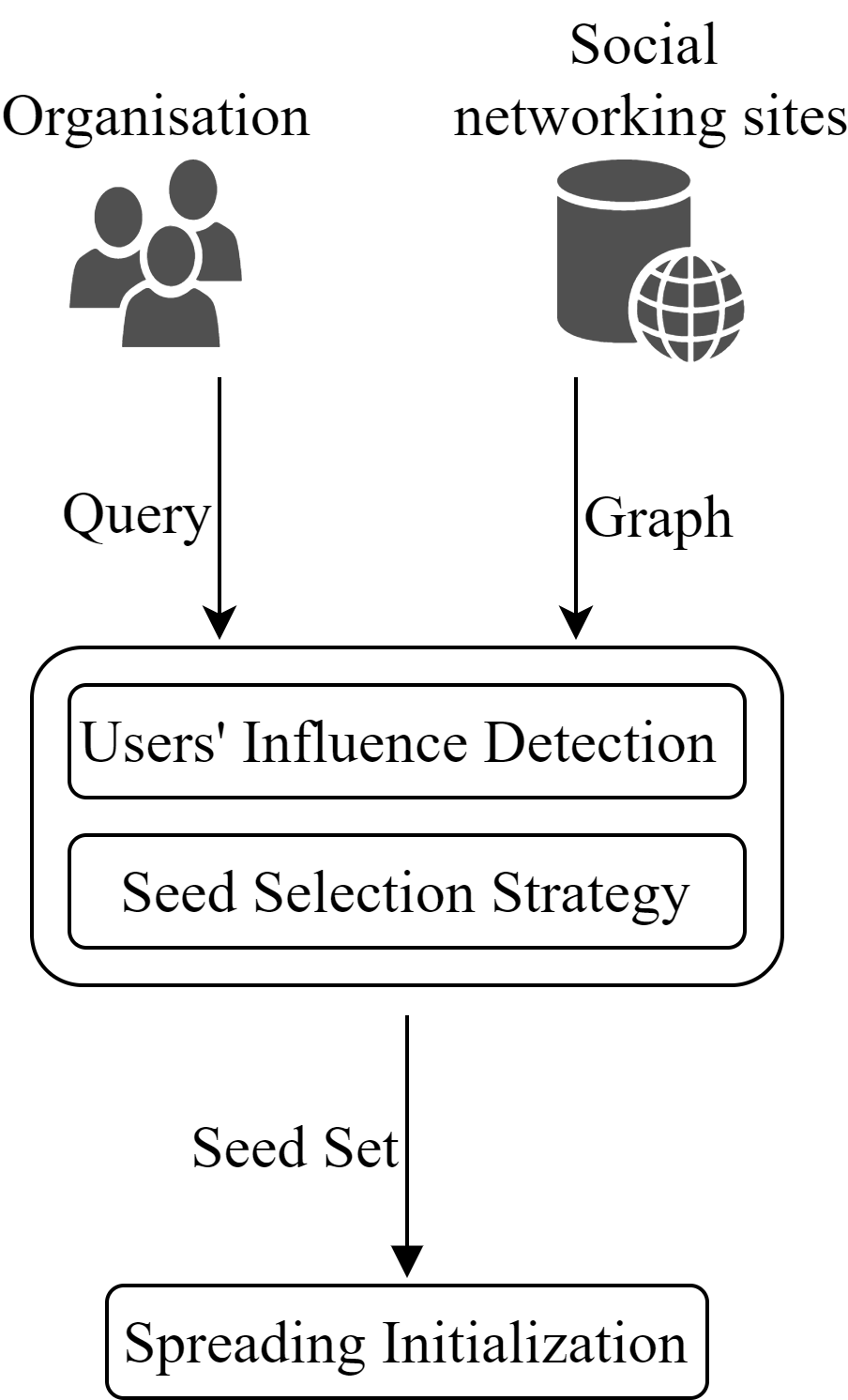}
\caption{Overview of the Influence Maximization problem.}
\label{Fig.modeling}
\end{center}
\end{figure}

Figure~\ref{Fig.modeling} gives an overview of the IM problem.
Somebody (say an organisation) intends to run a campaign to spread a message, advertisement, news or idea through a social network.
They formulate a query to identify a set of influential users that can help spread a specific message with specific features, preferences or constraints. For example, a company may consider advertising a new model of car for sale in a special exhibition; they may target users with an age greater than 18, who are located in the vicinity of the exhibition and they may have a limited budget. This can be formulated in the query as targets, geographical preference and budget constraint.
Social network sites typically provide data in relation to 
users and the relationships between them. Relevant user information may include shopping history, rate history, opinion, interests, geographical location and so on. Relationship may indicate some connection, such as friendship or common interests, between a pair of users. According to the information provided by social network sites and in line with the query formulated, influence maximization aims to identify a set of influential users as initial spreaders, known as a \emph{seed set}, to spread the message and maximize the total number of users who are influenced. The users who are influenced are called \emph{influenced users} or \emph{active users}.    
A formal model of the IM problem is given next.

First, an organisation formulates a query to define a message, its features and related information. The query is generally modelled as $Q=(M, AM, B)$. $M$ denotes the content of the message which may relate to news, an advertising slogan, an opinion, an event, and so on. $AM$ shows the different features of the message, such as topics of a product, location of an event and so on. $B$ may be the budget for propagation that can be the number of seeds, some discount for seeds to give an incentive to propagate a message, or remuneration to a network owner (or third-party service provider) to identify influential users and initiate a spreading process.

Second, given the features of the query, information provided by social network sites is used to model the social network; this step may be done by the owners of the network or by some third-party service providers. A social network is modelled as a graph $G=(V,E,BV,BE)$, with users and the relationships between them represented by nodes $V$ and edges $E$, respectively.  If there is a relationship between two nodes $v_i$ and $v_j$, it is shown by the edge $e_{ij}$; two nodes that are connected by an edge are considered as neighbours. $BV$ denotes the different behaviours of users such as location, interests, opinions and so on. $BE$ denotes the different features of the relationship between a pair of nodes indicated by the edges, such as trust, spread probability (also called as influence or activation probability), or common interests. Depending on the nature of relationships in the network, in some cases (for example, to represent the follow relationship in Twitter), the edges are considered as directed relationships, i.e. $e_{ij}\neq e_{ji}$, while in some other cases (for example, to represent friendship in Facebook) the relationship between nodes is regarded as bi-directional, and the graph is considered as undirected, i.e. $e_{ij}= e_{ji}$.  In general, a social network graph contains no cycles, i.e. $e_{ii}\not\in E$.

Third, according to the query $Q$ and graph $G$, the influence of each node is assessed and a strategy is applied to select a set of influential nodes as seeds taking into account the budget associated with the query and with an overall goal that the selected seeds maximize the influence of the spreading process. How to determine this set of influential nodes is the core of the influence maximization problem. Finally, the spreading process is initialized using the set of seeds to propagate the message.  

In what follows, an overview of the approaches applied in the literature to detect the influence of a user/a set of users is presented and the general framework of the influence maximization problem is formally defined.


\subsection {Influence Detection}

Different approaches have been described in the literature to detect the influence of a user or a set of users. This paper groups these approaches into the following four families.

\begin{itemize}
    \item Centrality measures determine the influence of each node in the social network graph based on topological properties. Different centrality measures have been proposed and extended to determine the influence of nodes~\cite{Degree,Betweenness,Closeness,ERM,H-index,ehc,K-shell}.
    
    \item Simulation of the spreading process can be used to determine the influence of each node and select a set of influential nodes. 
    Different diffusion models have been proposed to simulate the spreading process. 
    The Independent Cascade (IC) model~\cite{Cmodel1,Cmodel2,tarikh3} and the Linear Threshold (LT) model~\cite{TModel1,Tmodel2,tarikh3} are the models that have been widely applied.     
    In the IC model, the spreading process is \textit{simple}, which means the propagation on the edges is mutually independent and interaction with one active node may be enough for a node to be activated (influenced). In the LT model, the spreading process is \textit{complex}, which means a node may need to interact with multiple active nodes to be activated (influenced)~\cite{Surv8}. 
 
    \item Reverse Influence Sampling (RIS)~\cite{RIS1,RIS2,RIS3} relies on a random sampling technique to determine the influence of each node and identify a seed set. 
    
    \item Maximum Influence Arborescence (MIA)~\cite{MIA1} relies on the spreading probability on paths between pairs of nodes in the social network graph; the spreading probability on a path is calculated by multiplying the spreading probability on the edges of the path.  
    
\end{itemize}

Depending on the approach applied to detect the influence, we can approximately determine the time complexity of each method. Generally speaking, in terms of time complexity, these approaches can be ranked from high to low in the order: simulation-based, MIA-based, RIS-based and centrality-based.

\subsection{Influence Maximization}
The Influence Maximization (IM) problem aims to spread a message as widely as possible through a social network, taking into account that it is highly time consuming and practically impossible to send the message to all users of the network. As a result some constraints are taken into account to select a small set of influential users, who have more influence than other users and can widely propagate the message.
The IM problem is generally defined as in Eq.~(\ref{Eq.IM}), where the 
function $\varphi(S)$ defines the influence of a set of seeds $S$.  
\begin{equation}
    \begin{multlined}
        S^*=\underset{S\subset V}{arg\, max} \; \varphi(S)\\
        \mbox{subject to some constraints.}
    \end{multlined}
    \label{Eq.IM}
\end{equation}


Given the model discussed in Section~\ref{sec_PDescr}, there are three different types of features related to the behavioural information which may be taken into account to select a set of influential nodes: (i) features of the users ($BV$) to determine the relevance of each user to the query; (ii) features of the relationships ($BE$) to model the spreading (influence) probability between the users based on the relevance to the query; and (iii) features of the message ($AM$) to target a set of relevant users in spreading process. These three types of information encapsulate the differences between behaviour-agnostic and behaviour-aware methods. 

Behaviour-agnostic methods only consider the graph structure to determine the influence of nodes and identify a set of influential nodes. That is to say, behaviour-agnostic methods do not take into account anything specific for the sets $BV$ (capturing users' behaviour) and $BE$ (capturing edges' behaviour, that is, different features of the relationship between a pair of users). In addition, they do not consider different features of the message, as captured by the set $AM$ in the query message. In brief, behaviour-agnostic methods assume that: (i) all users have the same behaviour; (ii) relationships between any pairs of users are the same; and (iii) the features of the message have no impact on the propagation process. In behaviour-aware methods, some or all of these aspects may differ. 
It means that, in addition to graph structure, behaviour-aware methods take into account some aspects of behavioural information to propose practical approaches for real-world applications. 

In this paper, we focus on behaviour-aware methods; as mentioned in Section~\ref{Sec.Intro}, there are several surveys discussing behaviour-agnostic methods comprehensively. 


\section{Behaviour-Aware Methods}
\label{sec_beha_awr}
In behaviour-aware methods, besides graph structure, additional information, such as users' behaviour, relationships' features, and query content, is taken into account to improve the spreading process. For example, when a query aims to advertise a specific product, all users may not be equally interested in this product. Considering how interested each user is can improve the process of identifying a seed set and result in a successful message propagation. Based on the type of the information taken into account, behaviour-aware methods can be divided into four main categories.

\begin{figure}[ht!]
\begin{center}
\includegraphics[width=\textwidth]{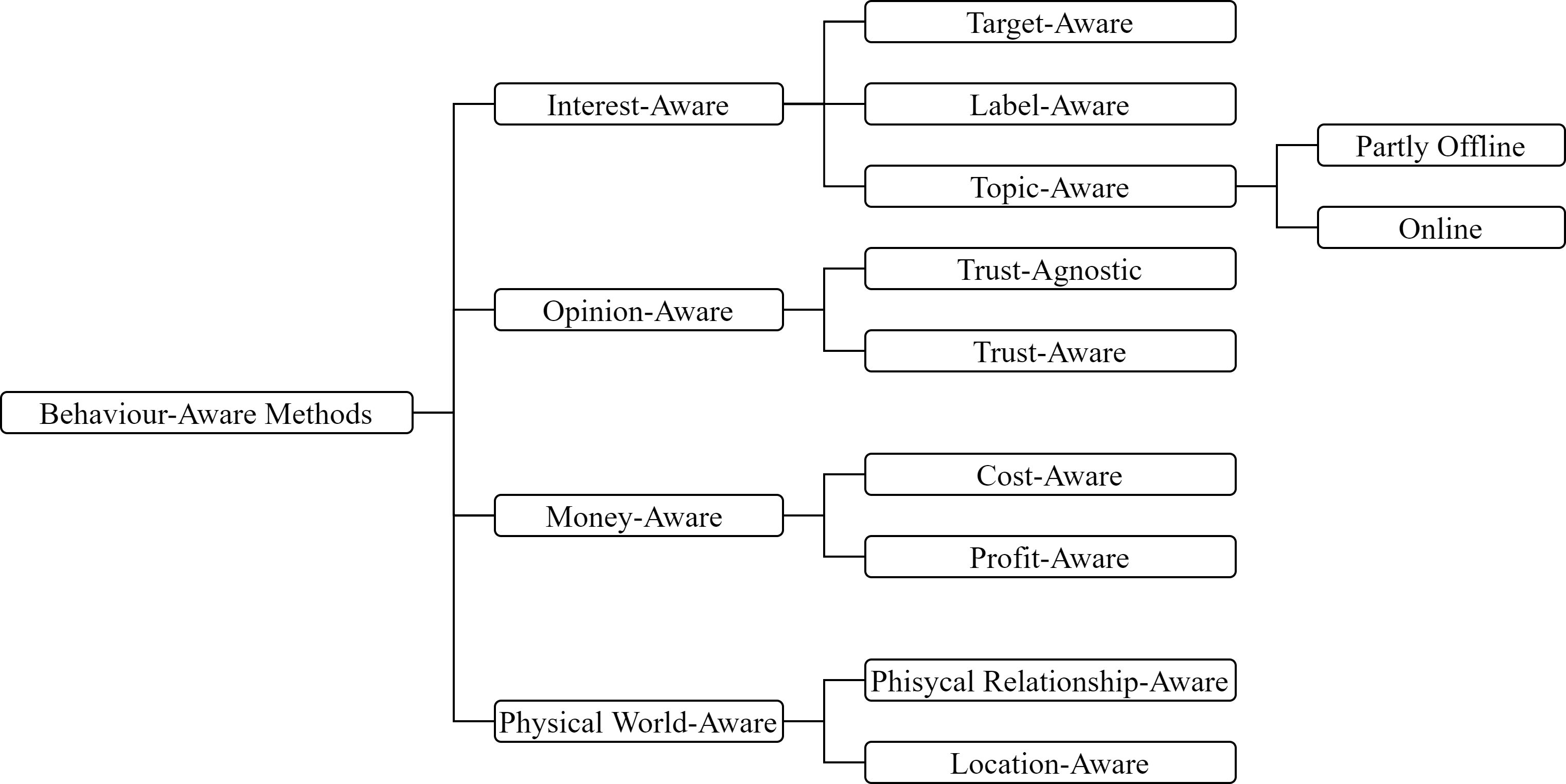}
\caption{A classification of behaviour-aware methods for IM}
\label{Fig.classification_behaviour_aware}
\end{center}
\end{figure}

User preferences and query relevance to these preferences are taken into account by \textit{interest-aware methods}. To determine user preferences, these methods rely on preprocessing of historical records of users' activities in the networks, such as, the content of their posts, the content of the posts they liked or replied to, their rating records for different products or their shopping experience and so on. In \textit{opinion-aware methods}, the opinion of users towards a message is taken into account to find influential users that can change the attitude or opinion of other users. To determine the stance and opinion of users towards a message, these methods rely on the analysis of the sentiment and subjectivity of users in their posts or their feedback on messages posted by other users. Cost and benefit of spreading an advertising query are considered by \textit{money-aware methods}. These methods assume that users may estimate their influence in the network and use it to negotiate a price to participate in the spreading process. Therefore, selecting different users as seeds may incur a different cost. Furthermore, the analysis of historical records of users' activities is leveraged in these methods to estimate users' valuation of a product advertised in a spreading process. This can help target users with a higher valuation and result in a higher benefit in the process. In \textit{physical world-aware methods}, the geographical location of users is taken into account and a query aims to spread a message towards users in specific locations, for example a query may invite users to a festival tacking place in a specific geographical location. These methods assume the location of users can be modelled using historical records of users' check-in locations, GPS-enabled technologies, location-based social networks, or similar techniques. These four categories are illustrated in Figure~\ref{Fig.classification_behaviour_aware} and will be further elaborated and discussed in detail in the sections that follow.

\subsection{Interest-Aware Methods}
Users' preferences and interests are taken into account in these methods. Spreading aims to influence the users who are potentially interested in a message. Past activities of users, such as posts, likes, ratings or shopping history may be preprocessed to determine their interests. In these methods, the size of the seed set is fixed, that is $|S|=k$, and it is also considered as a constraint of the IM problem as defined in Eq.~(\ref{Eq.IM}). We categorize these methods into three classes based on the process considered to identify the users interested in the message of the query formulated.

\subsubsection{Target-Aware Methods}
In target-aware methods, whether a user is interested in the message of the query or not is indicated by a simple binary value. In other words, the behaviour of users, corresponding to $BV$, is modelled with binary values to indicate whether a user is interested in a message or not. Users interested in a message are considered as target nodes. Edges and the query have no further information, i.e., the sets $BE$, and $AM$ are not relevant in these methods. IM aims to select a seed set to spread the message to as many target nodes as possible. 

In~\cite{m4,m5}, several centrality measures are extended to determine the influence of each node relative to the target nodes. In these measures, the centrality is calculated just based on the target nodes; other nodes have no impact in centrality calculation. Then, the top-k central nodes are chosen as seeds. In~\cite{m12,m11}, the influence of each node is determined based on the paths between the node and other nodes. In~\cite{m11}, two sets, \emph{influencers} and \emph{followers}, are defined for each node to determine the region of influence of the node. Then, the nodes are clustered based on their region; the influential nodes are greedily selected to maximize the activation probability of the target nodes in different regions. The authors in~\cite{m12} suggest a method based on the maximum influence arborescence method~\cite{MIA1}, which guarantees a solution within a factor of the optimal solution.

Propagation of the message to target nodes is also the objective in~\cite{m79,m18}. However, besides influence on target nodes, the diversity of the target nodes which are influenced is also taken into account. 
In~\cite{m79}, each target node is assumed to have a set of attributes; the problem is modelled as the identification of a seed set to spread the message to target nodes with diverse attributes. Then, a degree-based method is proposed to tackle the problem.
The authors in~\cite{m18} assert that some silent nodes may be interested in a message but may not have been considered as target nodes due to lack of past activity. Spreading the message to different parts of the network can motivate such nodes. Therefore, the authors apply the reverse influence sampling method~\cite{RIS1} to propose a method that selects a seed set to influence target nodes that have structural diversity. 

In~\cite{m80,m81}, the authors suggest that activating non-target nodes may bring undesired costs and results. Therefore, they discuss how to select a set of influential nodes to maximize the number of activated target nodes while simultaneously minimizing the number of non-target nodes activated. The authors in~\cite{m80} propose a simulation-based greedy method to identify a set of influential nodes such that the number of activated target nodes is maximized while the number of non-target activated nodes is kept below a given threshold; they propose heuristics for a time-efficient greedy method. In~\cite{m81}, maximizing the difference between the number of activated target nodes and activated non-target nodes is defined as the objective function of the problem. The authors propose a centrality measure to determine the influence of each node; influential nodes are selected in an iterative manner.  

In summary, considering a set of nodes as target nodes prevents waste of resources (in terms of time or money) and avoids propagation of a message towards nodes who may not be interested in the topic of a message. 
Additional information about the diversity of nodes, as for example in~\cite{m79,m18}, may boost the success of the spreading process. 
Taking into account non-target nodes, as in~\cite{m80,m81}, helps avoid the activation of nodes that may spread an adverse message and negatively impact the IM process.
However, relying on the diversity of nodes or non-target nodes needs information about users' behaviour which may not be always available. This suggests that the methods proposed in~\cite{m79,m18,m80,m81} are less viable than the methods in~\cite{m5,m4,m11,m12}.

\subsubsection{Label-Aware Methods}
Label-aware IM, also known as labelled IM, was first defined in~\cite{m1}. In label-aware IM, the features of a message, corresponding to $AM$ in the query formulated, are modelled using a set of labels relevant to the message. The labels are taken into account to model the spread of the message in the network.

In~\cite{m1,m13,m82}, a set of labels, such as health or news, is assigned to each node. Thus, the behaviour of each user, corresponding to $BV$, is modelled using a set of labels in which the user is interested. Target and influential nodes are selected, according to the overlap between label sets related to the query and behaviour of users. Two simulation-based methods are proposed in~\cite{m1} to identify influential nodes for label-aware IM; for time-efficiency, they also extend a degree-based method~\cite{SIMB1}. The authors in~\cite{m13} differentiate the state of the users between aware and adopted (in addition to inactive) and propose an extension of the linear threshold~(LT) model in which every node influenced by the propagation process moves to the aware state and forwards the message, but only interested nodes can switch to adopted state and accept the message. Then, a degree-based method~\cite{SIMB1} is developed to identify a seed set which maximizes the number of users in the adopted state. 
The impact of friend conformity in the propagation process is taken into account 
in~\cite{m82}. In order to take into account friend conformity, users are divided into groups based on the similarity of labels relevant to their behaviour; a centrality measure is then developed to determine the influence of each node and group. A set of seeds is selected based on the influence of nodes and groups.

In some studies~\cite{m23-1,m20}, instead of modelling the behaviour of each user, the features of the relationship between a pair of users are modelled using a set of labels; this set indicates the common interests between the pair based on their past communications. In~\cite{m20}, the spreading (influence) probability between each pair of nodes is determined based on the similarity of the label sets relevant to the query and the relationship between the pair. Then, a reverse sampling approach is proposed to identify the most influential seeds. In~\cite{m23-1}, the role of trust between users is also taken into account; the network is divided into domains based on the common labels and trust between users. Then, a degree-based method~\cite{SIMB1} is suggested to identify the influential users in each domain. 


In summary, in~\cite{m1,m13,m82}, the past activity of each user is used to determine a set of labels in which the user is interested. Taking into account this set may help identify influential seeds effectively and target nodes that are potentially interested. However, to determine the set of labels relevant to the messages exchanged between the pairs, considering past interactions between pairs of users, as in~\cite{m20,m23-1}, may help model the propagation of a message realistically. This is because two users may share some common interests but may not influence each other in all common interests; if past interactions were not considered, the spreading probability between them for some common interests may be zero.

\subsubsection{Topic-Aware Methods}

In these methods, a set of network topics is first defined. The behaviour of each user, corresponding to $BV$, can be modelled using a topic vector; the $j$-th entry of this vector is a number in $[0,1]$ and indicates how interested in the $j$-th topic the user is.  Also, the behaviour of each relationship, corresponding to $BE$, can be modelled using a topic vector; the $j$-th entry of this vector is a number in $[0,1]$ and indicates the spreading (influence) probability on the edge for the $j$-th topic. The features of the message, corresponding to $AM$ in the query, are modelled using a relevance topic vector whose $j$-th entry, a number in $[0,1]$, denotes the relevance of the query to the $j$-th topic. In topic-aware methods, influential nodes are selected according to these vectors. Based on the methodology applied to identify the seeds, we divide topic-aware methods into two categories.


\paragraph{Partly Offline Topic-Aware methods:}
In these methods, the main idea is that there are lots of different messages related to different sets of topics and the identification of influential users for each message is accordingly time-consuming. Therefore, different seed sets for different sets of topics are selected and saved in advance using an offline approach. When a query comes up, the overlap between the topic vector of the query and the topic sets for which seeds have been saved is used to determine the most relevant saved seed sets and select the most influential seeds.

The behaviour of each user is modelled using a topic vector in~\cite{m2,m17,m9}. In~\cite{m2}, reverse influence sampling~\cite{RIS1} is applied to determine and save offline the influence region of each node for different topics. When a query comes up, according to the influence region of nodes and the topic vector of the query, most influential nodes are identified as seeds. Some techniques are also proposed in this paper to enhance the time efficiency of the method. 
In~\cite{m17,m9}, the authors try to enhance the time efficiency of seed selection by dividing the topics into a set of groups based on the similarity of the topics. In~\cite{m17}, the authors argue that users with similar interests are attracted by messages with similar features. Thus, topics are divided into groups based on the similarity. A set of most influential nodes for each group is identified and saved offline. When a query comes up, a set of seeds is identified based on the similarity of the topic vector of the query and predefined influential nodes for topic groups. In~\cite{m9}, the authors observe that queries related to different topics may have a significant overlap; identifying an influential node set for each topic in advance and mixing the influential sets based on the query topics in online steps can speed up the seed selection. The authors adopt this hypothesis to propose a time-efficient method.


Instead of modelling the behaviour of users, in~\cite{m15} the features of the relationship between each pair of users is modelled using a topic vector, corresponding to $BE$, where the $j$-th value in the vector indicates the spreading (influence) probability of a message relevant to the $j$-th topic between the pair. Then, the maximum influence arborescence method~\cite{MIA1} is applied to develop a method to identify a set of seeds. Given the time complexity of this method, the authors apply an offline function to determine an upper bound for the influence of each node and reduce the computational complexity of seed selection.


\paragraph{Online Topic-Aware Methods:}
In online topic-aware methods, all steps to identify influential seeds are done online, when a query comes up; no preprocessing takes place. 

In~\cite{m10,m7}, the features of the relationship between each pair of nodes is modelled using a topic vector, corresponding to $BE$. In~\cite{m10}, according to the topic vector of the query, corresponding to $AM$, and the topic vector assigned to each edge, corresponding to $BE$, the spreading probability on each edge for the message is calculated. Then the IC and LT diffusion models are extended to identify a set of influential seeds. In~\cite{m7}, the authors take into account the role of communities in the network to select a set of influential seeds covering different parts of the network. They divide the network into a set of communities and select a number of influential seeds from each community according to the size of the community.  

In~\cite{m6,IMUD,m83,m84}, the behaviour of each user is modelled using a topic vector, corresponding to $BV$. In~\cite{m6}, the authors first discuss how to determine the interest of each user in different topics based on their activities in the network. Then, they propose a simulation-based method to identify a set of influential seeds based on the topic vectors of the query and users. In~\cite{IMUD}, the similarity between the topic vector of users and the topic vector of the query is calculated using entropy divergence notation to determine a weight for each node that indicates how interested the node is in the query's message. According to the nodes' weight, a weight is then assigned to each edge, and the influential nodes are selected based on the weight of the edges connected to each node. 
In~\cite{m83}, a network embedding approach is proposed to capture the interests of users; then, this approach is applied to determine the influence of each node. A set of influential nodes is identified using a reinforcement learning-based algorithm. In~\cite{m84}, the authors argue that greater similarity between the interests of a user and features of the query is an indication that the user may be influenced in the spreading process. Thus, they calculate the similarity between the topic vector of users and the topic vector of the query to determine the probability that each node is influenced in the spreading process. Then, an independent cascade model is extended to capture this property; a degree-based heuristic is proposed to determine the influence of each node and select the seeds in an iterative manner. 


In summary, in some methods (partly offline methods~\cite{m2,m17,m9} and online methods~\cite{m6,IMUD,m83,m84}), the behaviour of users is taken into account to determine how interested each user is in the query's message. On the other hand, in other methods (partly offline methods~\cite{m15} and online methods~\cite{m10,m7}), the log of the interactions between each pair of users is used to model the  features of the relationship between the pair. As discussed before, taking into account the features of relationships may help  model the propagation process in the network  effectively. To compare partly offline and online methods, when a query comes up, partly offline methods may identify influential nodes to trigger the spreading process more efficiently, which is an advantage especially for breaking news spreading in a social network. However, partly offline methods need to deal with the running time for offline execution and also storage issues in relation to saving the influential seeds for queries with different topic vectors. Another challenge of partly offline methods is how to determine the potential queries in advance to identify the influential seeds for different topic vectors. 

\subsubsection{Summary}

The properties of the discussed interest-aware methods are summarized in Table \ref{Tabel.Interest-based}. As mentioned, in target-aware methods, a binary value is used for each user to determine whether the user is interested in the query's message or not, something that may not properly capture the views of a user towards a message. Label-aware methods try to address this by taking into account a set of labels in which the user is interested, these methods also use a binary state to represent whether a user is interested in a special topic or not. Conversely, topic-aware methods capture user preferences more realistically by applying a continuous value to represent the interest of each user towards a topic. However, determining the topic vector for each user demands a good record of historical data of users' past activities which may not be always available.

\begin{table*}[ht!]
\centering\footnotesize
\caption{Properties of interest-aware methods, including applied behavioural features, applied method for influence detection -- Centrality Measure (CM), Spreading Simulation (SS), Reverse Influence Sampling (RIS), and Maximum Influence Arborescence  (MIA) -- and the type of spreading process}
\resizebox{\textwidth}{!}
{
\begin{tabular}{|c|c||c|c|c||c|c|c|c||c|c|}
 \hline
\multirow{2}{4em}{Category} & \multirow{2}{4em}{\centering Paper (year)} & \multicolumn{3}{c||}{Behavioural features} & \multicolumn{4}{c||}{Influence detection}&\multicolumn{2}{c|}{Spreading process}\\ 
\cline{3-11}
&&AM&BV&BE&CM & SS & RIS & MIA&Simple&Complex\\
 \hline
\multirow{5}{4em}{Target-Aware}& \cite{m5}(2003) &&\checkmark&&\checkmark&  & & &\checkmark &\\
\cline{2-11}
& \cite{m4}(2014) & &  \checkmark&& \checkmark& & & &\checkmark& \\
\cline{2-11}
& \cite{m11}(2018) & &\checkmark&& & & &\checkmark&\checkmark&\\
\cline{2-11}
& \cite{m12}(2014) & &\checkmark&&& & &\checkmark&\checkmark&\\
\cline{2-11}
& \cite{m18}(2018) & &\checkmark&&&&\checkmark& &&\checkmark \\
\cline{2-11}
& \cite{m79}(2021) & &\checkmark& && &\checkmark& &\checkmark&\\
\cline{2-11}
& \cite{m80}(2018) & &\checkmark & &&\checkmark& & &\checkmark&\checkmark\\
\cline{2-11}
& \cite{m81}(2020) &\checkmark&&\checkmark&& & &  &\checkmark& \\
\hline

\multirow{4}{4em}{Label-Aware}& \cite{m1}(2011) & \checkmark &\checkmark&&\checkmark&\checkmark & & &&\checkmark \\
\cline{2-11}
& \cite{m23-1}(2015) & \checkmark &&\checkmark&\checkmark& &  &  &\checkmark&\\
\cline{2-11}
& \cite{m82}(2018) &\checkmark&\checkmark&\checkmark&\checkmark& &  &  &\checkmark& \\
\cline{2-11}
& \cite{m13}(2017) & \checkmark & &\checkmark&\checkmark&  &  &  &&\checkmark \\
\cline{2-11}
& \cite{m20}(2018) &\checkmark  & &\checkmark& &  &\checkmark&&\checkmark& \\

\hline
\multirow{4}{4em}{Partly Offline Topic-Aware}& \cite{m2}(2015)&\checkmark & \checkmark&  & &  &\checkmark&&\checkmark&\checkmark \\
\cline{2-11}
&\cite{m17}(2014) & \checkmark  &    & \checkmark  &&\checkmark&  &  &\checkmark& \\
\cline{2-11}
&\cite{m9}(2016) & \checkmark&&\checkmark&  &\checkmark& &  &\checkmark& \\
\cline{2-11}
&\cite{m15}(2015) &\checkmark&&\checkmark& &  &  &\checkmark&&\checkmark\\

\hline
\multirow{4}{4em}{Online Topic-Aware}& \cite{m10}(2013) &\checkmark&\checkmark&&&\checkmark &  &   &\checkmark&\checkmark  \\
\cline{2-11}
& \cite{m7}(2019) &  \checkmark &&\checkmark&& \checkmark &  &  &\checkmark&\checkmark  \\
\cline{2-11}
& \cite{IMUD}(2019) &\checkmark&\checkmark&&\checkmark&& &  &  \checkmark& \\
\cline{2-11}
& \cite{m6}(2014)  & \checkmark  & \checkmark  &  &  & \checkmark & & &\checkmark&  \\
\cline{2-11}
& \cite{m83}(2020) & \checkmark&\checkmark&  &\checkmark &  &  &  &\checkmark &\checkmark  \\
\cline{2-11}
& \cite{m84}(2020)  & \checkmark& \checkmark& & \checkmark&  & &  & \checkmark&  \\
\hline
\end{tabular}
}
\label{Tabel.Interest-based}
\end{table*}

\subsection{Opinion-Aware Methods}
In these methods, each user has an opinion or attitude about the query's message, i.e., the behaviour of each user towards the query's message is modelled using a discrete or a continuous value. Using a discrete value, opinion has a limited set of choices and can often be indicated by a binary choice between positive/negative or a choice between opinions A/B; in this case the goal of IM is to maximize the number of users subscribing to a specific opinion. Using a continuous value, the opinion of each user is denoted by a real value to express the leaning of users towards the query's message; in this case IM is about selecting a set of influential users that can potentially maximize this value for the network users.


It has been often mentioned that opinion change depends on the trust between the users. Thus, it has been proposed to model social networks as a signed graph whose edges have a positive or negative sign indicating trust or distrust between users. Yet, trust between the users is not taken into account by all methods. Thus, it is useful to divide opinion-aware methods into two categories: trust-agnostic and trust-aware. Same as with interest-aware methods, in opinion-aware methods, $|S|=k$ is also considered as a constraint of the IM problem as defined in Eq.~(\ref{Eq.IM}), which means the size of the seed set  is fixed.


\subsubsection{Trust-Agnostic Methods}
In~\cite{m48,m43,m45}, the opinion of each user is modelled using a continuous value; a higher value corresponds to a more favorable opinion towards the query's message. In~\cite{m43}, the authors develop an opinion cascade model to simulate the spreading process and how users' attitude may change over this process. Then, this model is applied to suggest a greedy method to find a seed set, which, when triggered, maximizes the total opinion of influenced users. In~\cite{m48}, the behaviour of each user is modelled using two opinions: internal opinion, which persists over time, and expressed opinion, which is influenced by other users over time. Taking into account both internal and external opinions, the changing of expressed opinion of users is modelled; a greedy method is suggested which greedily selects the seeds in an iterative manner. The authors in~\cite{m45}, in addition to the opinion of users, model the features of relationship between the users by taking into account both spreading probability and interaction probability on the edges. Inspired from fluid dynamics, a diffusion model is proposed to simulate the change of users' opinion over spreading process. This model is then used to describe a greedy method to find a seed set which, when triggered, maximizes the number of active users whose opinion is greater than a given threshold.

In~\cite{m85,m86}, the authors model the behaviour of each user using a continuous value; this value is considered as a threshold to determine whether the user is influenced during the spreading process. The authors assert that traditional diffusion models (independent cascade and linear threshold models) cannot properly define the dynamics of users' opinion. In~\cite{m85}, the independent cascade model is used to simulate the spreading process, while the dynamics of the users' opinion is captured using a voter model; combining these two models a multi-stage diffusion model is proposed. The influence of each node is determined using a centrality measure; a greedy method along with a heuristic algorithm is suggested to identify a set of seeds. In~\cite{m86}, the authors argue that, because users' opinion changes dynamically over  time, seeds should be selected in an adaptive manner to guarantee the quality of the spreading process. They apply reinforcement learning theory to propose a multi-stage heuristic to identify influential seeds. In this method, the spreading process is initialized with a batch of seeds; additional seeds are selected during the spreading process.

In~\cite{m39,m41}, the opinion of each user is modelled using a discrete value (negative, neutral, or positive) that models attitude towards the query's message; IM aims to maximize the number of users with positive opinion. In~\cite{m39}, a cascade model is proposed to simulate the spreading process with negative opinions. The authors first apply this model to propose a simulation-based method; then, maximum influence arborescence~\cite{MIA1} is applied to propose a time-efficient method. In~\cite{m41}, the authors argue that when spreading a product advertisement users' complaining behaviour should be taken into account; a user may be satisfied with a product but may still have something to complain about. Thus, in addition to users' opinion, they also model the complaining behaviour of users during the spreading process, which is considered by a simulation-based method to identify an influential seed set.

In summary, in~\cite{m39,m34}, the opinion of each user is modelled using a discrete value 
while, in~\cite{m48,m43,m45,m85,m86}, a continuous value is used to model the opinion of each user. Compared to the methods in~\cite{m48,m43,m45}, the proposed methods in~\cite{m85,m86}, try to model more accurately how the opinion of users changes under the influence of friends. However, the dynamics of users' opinion under the influence of their friends may not be straightforward and cannot be modelled for all users in the same way.

\subsubsection{Trust-Aware Methods}
Users may consider other users as either friends or foes~\cite{m46}; thus, considering both positive and negative relationships can improve opinion-aware methods. In trust-aware methods, each user has their own opinion towards the query's message; the opinion of users may change under the influence of friend or foe relationships with other users. The friend or foe relationship is typically represented by a sign; edges with a positive sign indicate friendship between users, while a negative sign indicates a foe relationship. Therefore, in trust-aware methods, the feature of the relationship between a pair of nodes is modelled using a sign describing the friend or foe relationship.  


In some methods~\cite{m28,m30,m40,m29}, the behaviour of each user is modelled as a continuous value, falling in $[-1,1]$, to indicate the opinion of the user about the query's message; the behaviour of each relationship is modelled as a sign indicating friend or foe relationship. In~\cite{m28}, the authors present an extension of the linear threshold~(LT) model to simulate propagation in a signed network and update users' attitude over propagation. Applying this model, a simulation-based method is proposed to select a set of seeds that maximize the number of nodes with a positive opinion. In order to propose a time-efficient method for this problem, in~\cite{m30}, a centrality measure is defined to determine the influence of each node based on the friend and foe relationships of the node in a three-hop neighbourhood; nodes with the greatest centrality are selected as seeds. In~\cite{m40}, the authors extend the linear threshold model to take into account different states for each user during the spreading process; how the state of each user changes is determined using some given thresholds. The authors suggest a simulation-based greedy algorithm to identify an influential seed set. In~\cite{m29}, the behaviour of each user is modelled using two continuous values: internal opinion, which does not change during the spreading process, and expressed opinion, which may be influenced and change during the propagation as a result of the influence of the other users. The authors develop a linear threshold model to simulate the propagation in this setting and propose a greedy method along with some heuristics to select a set of influential nodes.

In some other methods, the opinion of each user towards the query's message is modelled as a discrete value, such as negative/neutral/positive, A/B, or red/blue. Without loss of generality, we can refer to all these as negative, neutral or positive opinion. Here, the aim of IM is to maximize the number of users with positive opinion after propagation.

In~\cite{m36,m25,m44,m87,m42,m37}, the independent cascade model is adopted as the diffusion model to simulate the spreading process in the network. In~\cite{m36} a diffusion model, named polarity-related independent cascade, is developed to simulate propagation in a signed network and change of users' opinion. Applying this model, a polarity-related method is proposed to find a set of nodes with positive opinion which, when triggered, maximize the number of nodes with positive opinion. Due to the low time-efficiency of this method, some research has tried to improve the time complexity of seed selection under the polarity-related independent cascade model; the authors in~\cite{m25,m44} apply meta-heuristic algorithms, while the authors in~\cite{m87} propose a path-based centrality measure to identify influential nodes. In~\cite{m42}, the authors extend the problem by assuming that nodes with negative opinion can also be selected as seeds to initiate competitive propagation in the network; a centrality measure is suggested to address the problem. In~\cite{m37}, it is assumed that when a node is influenced by one of its neighbours, it takes some time until the node attempts to influence other neighbours; therefore, in addition to the trust, time latency is defined to model the behaviour of each relationship. The authors develop an extension of the independent cascade model to take into account both trust and time latency in the spreading process; then, a greedy algorithm, with some heuristics for efficiency improvement, is proposed to identify an influential seed set under the extended model.

In~\cite{m38,m31,m88}, the linear threshold model is adopted as the diffusion model. In~\cite{m38}, the authors propose an extension of the linear threshold model to simulate the propagation process and dynamics of users' opinion under the influence of neighbours with opposing opinions. A method is then proposed which greedily selects seeds using a simulation-based approach. In line with the diffusion model presented in~\cite{m38} but using a different model for how the opinion of users changes during the spreading process, an extension of the linear threshold model is proposed in~\cite{m31,m88}; these papers apply some heuristics to efficiently identify a set of influential nodes. In~\cite{m31}, a centrality measure under the extended linear threshold model is defined to detect the influence of each node; a set of influential nodes is selected using a heuristic method.
The authors in~\cite{m88} identify a set of influential nodes using a simulation-based method incorporated with two acceleration techniques: (i) remove a set of nodes with small spreading ability, and (ii) prune the number of paths to reduce the time to simulate the spreading process.

Some studies~\cite{m46,m34,m33,m89,m90} avoid conventional diffusion models when  simulating the spreading process in order to take into account the formation of users' opinion under the influence of neighbours. A voter model is proposed in~\cite{m46} to analyse the influence diffusion dynamics mathematically. The authors apply this model to determine the influence of nodes in long- and short-term propagation. This voter model is employed in~\cite{m34} to develop an extension of pagerank centrality~\cite{PageRank}, which determines the influence of nodes. The impact of community structure on the influence of a set in signed networks is investigated in~\cite{m33}. For this purpose, the spreading process is simulated using a game-theoretic approach to assess the correlation between the influence of different seed sets and inter- and intra-community edges. The authors conclude that there is a significant correlation between the influence ability of a seed set and community structure.
In~\cite{m89}, the authors argue that the linear diffusion model may not appropriately determine the dynamics of users' opinion in the spreading process; the authors propose an extended linear threshold model that incorporates an opinion formation model. They also propose a centrality measure to determine the influence of each node in this model; a greedy method along with a heuristic algorithm is suggested to identify a set of seeds. In~\cite{m90}, reinforcement learning is used to model how the users' opinion changes during the spreading process and propose an adaptive seed selection method where some seeds are selected during the spreading process.

In summary, modelling the opinion of users using continuous values~\cite{m28,m30,m40,m29} may capture their opinion more accurately compared to discrete values~\cite{m36,m25,m44,m87,m42,m37,m38,m31,m88,m46,m34,m33,m89,m90}. 
The methods proposed in~\cite{m46,m34,m33,m89,m90} try to model the dynamics of users' opinions differently from the spreading process. In fact, opinion dynamics is a topic that may need further attention in future research.

\subsubsection{Summary}

The properties of the discussed opinion-aware methods are summarized in Table~\ref{Tabel.opinion-maximization}. Trust-agnostic methods consider that all network relationships are positive (friend), whereas in reality networks may also include negative relationships (foe); user reaction to suggestions may differ depending on whether it comes from a friend or a foe. Overall, there is scope to model positive and negative relationships more elaborately than what has been attempted in the research literature up to now.


\begin{table*}
\centering\footnotesize
\caption{Properties of opinion-aware methods, including applied behavioural features, applied method for influence detection -- Centrality Measure (CM), Spreading Simulation (SS), Reverse Influence Sampling (RIS), and Maximum Influence Arborescence  (MIA) -- and the type of spreading process}
\resizebox{\textwidth}{!}
{
\begin{tabular}{|c|c||c|c|c||c|c|c|c||c|c|}
 \hline
\multirow{2}{4em}{Category} & \multirow{2}{4em}{\centering Paper (year)} & \multicolumn{3}{c||}{Behavioural features} & \multicolumn{4}{c||}{Influence detection}&\multicolumn{2}{c|}{Spreading process}\\ 
\cline{3-11}
&&AM&BV&BE& CM & SS & RIS & MIA&Simple&Complex\\
 \hline
 \multirow{5}{4em}{Trust-agnostic}& \cite{m48}(2013) &&\checkmark&& \checkmark & &&&&\checkmark\\
\cline{2-11}
& \cite{m43}(2013)&&\checkmark&& &\checkmark & &&&\checkmark\\
\cline{2-11}
& \cite{m45}(2018)&&\checkmark&&&\checkmark& &&\checkmark&\checkmark \\
\cline{2-11}
& \cite{m85}(2021)&&\checkmark&&\checkmark&& &&&\checkmark\\
\cline{2-11}
& \cite{m86}(2021)&&\checkmark&&&\checkmark&&&\checkmark& \\
\cline{2-11}
& \cite{m39}(2011)&& \checkmark&&\checkmark&&&\checkmark&\checkmark&\\
\cline{2-11}
& \cite{m41}(2012)&& \checkmark&&\checkmark&\checkmark&&&\checkmark&\\
\hline

\multirow{13}{4em}{Trust-Aware}&\cite{m46}(2013) &&\checkmark&\checkmark&\checkmark&&&&&\checkmark\\
\cline{2-11}
& \cite{m28}(2015)&&\checkmark&\checkmark&&\checkmark&&&&\checkmark \\
\cline{2-11}
& \cite{m29}(2019)&& \checkmark&\checkmark&&\checkmark& &&& \checkmark \\
\cline{2-11}
& \cite{m30}(2016)&&\checkmark&\checkmark&\checkmark&& &&&\checkmark \\
\cline{2-11}
& \cite{m34}(2015)&&\checkmark&\checkmark&\checkmark& &&&&\checkmark\\
\cline{2-11}
& \cite{m36}(2014)&&\checkmark&\checkmark&&\checkmark&&&\checkmark& \\
\cline{2-11}
& \cite{m25}(2017)&&\checkmark&\checkmark&\checkmark&&&&\checkmark& \\
\cline{2-11}
& \cite{m87}(2020)&&\checkmark&\checkmark&\checkmark&&&&\checkmark& \\
\cline{2-11}
& \cite{m42}(2015)&& \checkmark &\checkmark&\checkmark&&&&\checkmark& \\
\cline{2-11}
& \cite{m44}(2016)&&\checkmark & \checkmark&&\checkmark&&&\checkmark&\\
\cline{2-11}
& \cite{m37}(2015)&&\checkmark&\checkmark&\checkmark&\checkmark&&&\checkmark& \\
\cline{2-11}
& \cite{m31}(2016)&&\checkmark&\checkmark&\checkmark&&&&&\checkmark\\
\cline{2-11}
& \cite{m88}(2019)&&\checkmark&\checkmark&&\checkmark&&&&\checkmark\\
\cline{2-11}
& \cite{m38}(2013)&&\checkmark&\checkmark&&\checkmark&&&&\checkmark\\
\cline{2-11}
& \cite{m40}(2016)&&\checkmark&\checkmark&&\checkmark&&&&\checkmark\\
\cline{2-11}
& \cite{m89}(2021)&&\checkmark&\checkmark&&\checkmark&&&&\checkmark\\
\cline{2-11}
& \cite{m90}(2022)&&\checkmark&\checkmark&\checkmark&&&&&\checkmark\\
\hline
\end{tabular}
}
\label{Tabel.opinion-maximization}
\end{table*}


\subsection{Money-Aware Methods}

In these methods monetary aspects of the spreading process are taken into account. Some money (or equivalent, such as discount on a product) is paid to users to persuade them to take part in the spreading process as a seed. Alternatively, the activation of different users may have different monetary benefits in the spreading process. Overall, money-aware methods can be divided into two categories: cost-aware methods, where a budget constraint is considered in seed set selection, and profit-aware methods where, in addition to budget, specific decisions during the spreading process may lead to monetary benefits.

\subsubsection{Cost-Aware Methods}
In these methods, selection of each user as a seed is accompanied with a cost; the behaviour of each user, corresponding to $BV$, is modelled as a value indicating the cost that the user may ask in order to agree to take part in the spreading process as a seed. In this case, the constraint in the query, corresponding to $B$, is the total budget; the sum of the costs of the seeds must not exceed this budget. This implies that, in these methods, instead of specifying a fixed number of seeds, as it was the case with the methods discussed earlier, the query is driven by a budget to identify a seed set. 

Influence maximization using a budget constraint, also known as budgeted-IM, was first introduced 
in~\cite{m51}. The authors defined the problem under the independent cascade model; this model was applied to propose a greedy method to find seeds in several iterations. In each iteration, the node with the greatest influence to cost ratio is added to the seed set. They also applied the maximum influence arborescence method~\cite{MIA1} to propose a time-efficient greedy algorithm to solve the problem. In~\cite{m59}, the authors evaluate the impact of considering the cost of nodes and the budget constraint under the independent cascade model. They assess four well-known centrality measures to select top nodes as seeds and show that these measures cannot effectively identify a highly influential seed set in cost-aware IM.


In~\cite{m63,m50,m53}, the authors propose time-efficient methods to identify influential nodes under the independent cascade model. Two iterative functions are proposed in~\cite{m63} to determine the influence, susceptibility of being influenced and the persuasion cost of each node based on the properties of the node. Then, a method is proposed to select seeds iteratively; in each iteration, based on the cost and influence of nodes, a seed is selected and the susceptibility of being influenced of the other nodes is updated. In~\cite{m50}, using a centrality measure, a set of nodes is selected as candidate seeds; the set contains the nodes with greatest influence and/or least cost. The authors apply a meta-heuristic algorithm to develop a combination strategy to identify the best seeds as a composition of the nodes with great influence and low cost. In~\cite{m53}, the problem is modelled using binary integer programming. The author first applies the independent cascade model to propose a simulation-based method to identify influential seeds; then, a sampling approach is developed to propose a time-efficient method which guarantees an approximation to the optimal solution.

In~\cite{m58}, the problem is defined under the linear threshold model. This model is used to determine a surrounding set for each node which contains the nodes with the least cost to activate the node. This set is used to determine the influence to cost ratio of the node; then, a set of influential nodes is identified based on this ratio. In~\cite{m57,m56}, the problem is solved using knapsack constraints. In~\cite{m57}, the authors apply the history of the relationship of each user with its direct and indirect neighbours to determine the influence of the user. Seed set selection is described as an iterative process; in each iteration, the node with the greatest influence is added to the seed set if it satisfies a particular cost based criterion. The problem is extended in~\cite{m56} to propagate multiple messages about multiple items. Some constraints on items are first considered and the problem is modelled using knapsack constraints. Then, based on the cost-efficiency of assigning a specific item to each user, an adaptive method is proposed to identify influential seeds for each item. 

The role of community structure in propagation has been taken into account in~\cite{m54,m93}. For this purpose, in~\cite{m54}, the network is first divided into communities and a portion of the budget is assigned to each community, based on the number of nodes in the community. In each community, a set of nodes is iteratively selected as seeds; in each iteration the node with the maximum degree whose cost is not greater than the remaining budget of the community is selected as a seed. 
In~\cite{m93}, the authors take into account the interests of users in each community to determine each user's influence. They assume the network contains a set of topics; users in each community are interested in a subset of these topics. Then, applying maximum influence arborescence~\cite{MIA1}, a method is proposed to identify influential users and influential topics in each community.

In~\cite{m94}, the authors address the problem from a different point of view. They consider a set of advertisers, each of whom has a limited budget and aims to spread a message to advertise a product. Given each advertiser's budget, the owner of the network selects a seed set for each advertiser to initialize the spreading process. According to the costs of seeds and also the number of influenced users for the message corresponding to each advertiser, the benefit of the owner is determined. The question is how to select a set of influential users for each advertiser so that the total benefit of the owner of the network is maximized. To tackle the problem, the authors propose sampling methods to identify the influential users that meet this goal.

\subsubsection{Profit-Aware Methods}

Apart from monetary incentives to persuade users to act as seeds in the spreading process, the activation of each user may bring some benefit for the organisation running the campaign. In profit-aware methods, in addition to the cost, a monetary benefit is defined to model the behaviour of each user; the benefit is a value which is gained if the user is activated during the spreading process. The goal is to maximize the overall profit based on the cost and benefit of the spreading process.


In~\cite{m16,m62,m61}, the behaviour of each user is modelled using two values indicating its cost if selected as a seed, and its benefit if activated in the propagation process. In~\cite{m16}, the IM problem is defined as identification of a seed set whose cost is not greater than a given budget and maximizes the total benefit from all activated nodes during the spreading process. The reverse influence sampling method~\cite{RIS1} is then used to find influential users; nodes that have greater influence on nodes returning high benefits are considered as influential nodes. In \cite{m62,m61}, the problem is defined as the identification of a set of nodes to maximize profit, i.e., benefit minus cost, without any budget limitations. In~\cite{m62}, the authors first propose a greedy method to solve the problem; they further propose a double greedy method to propose a method which guarantees an approximation of the optimal solution of the problem. The problem is further extended in~\cite{m61} by defining two features (price of a product offered to each user and valuation of the user) to model the decision-making behaviour of each user. Then, in order to simulate the spreading process, the linear threshold model is extended to incorporate these features. This model is applied to propose a greedy method to identify the seeds that maximize the profit.   


In~\cite{m91,m92}, the authors assume that influencing the users who are not interested in the query's message brings no gain; therefore, a set of interested users is considered as the target of the spreading process. According to the distance (hops) between a pair of nodes, a heuristic method is proposed in~\cite{m91} to identify a set of influential nodes maximizing the total benefit. In order to improve this process, the role of communities is taken into account in~\cite{m92}. In this method, based on the number of target nodes in each community, a proportion of the budget is assigned to the community; a centrality measure is then proposed to identify influential users in each community.

Profit-aware IM is further extended in~\cite{m60,m64}. Instead of activating nodes, in~\cite{m60}, the aim is activating communities. It is assumed that the activation of each community brings a benefit; a community is activated if at least a given percentage of the nodes in the community is activated. Then, the IM problem is defined as identifying a seed set for which the profit, i.e., the total benefit of the activated community minus the total cost of the seeds, is maximized. The authors apply the reverse influence sampling method~\cite{RIS1,RIS2} to propose a method which guarantees an approximation of the optimal solution. Profit-aware IM is extended in~\cite{m64} by considering the query formulated to advertise a set of different products; the behaviour of each user towards the adoption of each product is modelled by an adoption threshold; the adoption of each product by the user brings different benefits. The linear threshold model is extended to incorporate these features. The authors first apply this model to propose an iterative greedy method to identify the seeds that maximize profit; in each iteration, the node with the greatest influence to cost ratio is added to the seed set as a new seed. Then, some heuristics are developed to suggest a time-efficient method. The authors also model the problem as a multi-choice knapsack problem to distribute the budget across the products optimally; then, they apply binary integer programming to solve the problem.     


\subsubsection{Summary}

The properties of the discussed money-aware methods are summarized in Table~\ref{Tabel.Monetary-aware}.
Cost-aware methods try to model IM in social networks by taking into account the costs of the spreading process. However, these methods assume that all influenced users bring the same benefit. On the other hand, profit-aware methods try to model the problem by taking into account monetary concerns of users in real-world scenarios. For example, in online advertisements, people have their own valuations for buying the product advertised by a campaign. Therefore, influencing different users may bring different benefits to the campaign; considering this aspect of the problem helps profit-aware methods boost the success of the spreading process compared to cost-aware methods. It is noted that data about users' monetary concerns and valuations may not be always readily available, which means that cost-aware methods may be more relevant than profit-aware methods in practice.   

\begin{table*}
\centering\footnotesize
\caption{Properties of money-aware methods, including applied behavioural features, applied method for influence detection -- Centrality Measure (CM), Spreading Simulation (SS), Reverse Influence Sampling (RIS), and Maximum Influence Arborescence  (MIA) -- and the type of spreading process}
\resizebox{\textwidth}{!}
{
\begin{tabular}{|c|c||c|c|c||c|c|c|c||c|c|}
 \hline
\multirow{2}{4em}{Category} & \multirow{2}{4em}{\centering Paper (year)} & \multicolumn{3}{c||}{Behavioural features} & \multicolumn{4}{c||}{Influence detection}&\multicolumn{2}{c|}{Spreading process}\\ 
\cline{3-11}
&&AM&BV&BE& CM & SS & RIS & MIA&Simple&Complex\\
 \hline
 \multirow{5}{4em}{Cost-Aware}& \cite{m51}(2013) &&\checkmark&&&\checkmark&&&\checkmark& \\
\cline{2-11}
& \cite{m50}(2014)  &&\checkmark&&\checkmark&&&&\checkmark& \\
\cline{2-11}
&  \cite{m54}(2019) &&\checkmark&&\checkmark&&&&\checkmark& \\
\cline{2-11}
&  \cite{m93}(2020) &&\checkmark&\checkmark&&&&\checkmark&\checkmark&  \\
\cline{2-11}
&  \cite{m57}(2018) &&\checkmark&&\checkmark&&&&\checkmark&  \\
\cline{2-11}
& \cite{m56}(2013) &\checkmark&\checkmark&&\checkmark&&&&\checkmark&  \\
\cline{2-11}
&  \cite{m53}(2019)&&\checkmark&&&\checkmark&\checkmark&&\checkmark&  \\
\cline{2-11}
&  \cite{m63}(2015)&&\checkmark&&\checkmark&&&&\checkmark&  \\
\cline{2-11}
&  \cite{m58}(2020) &&\checkmark&& \checkmark &&&&&\checkmark\\
\cline{2-11}
&  \cite{m94}(2021) && \checkmark&&&& \checkmark&& \checkmark &  \\
\hline

\multirow{5}{4em}{Profit-Aware}&\cite{m16}(2016) &&\checkmark&&&&\checkmark&&\checkmark&\checkmark\\
\cline{2-11}
&\cite{m64}(2016)&\checkmark&\checkmark&&\checkmark&\checkmark&&&&\checkmark\\
\cline{2-11}
&\cite{m61}(2012)&\checkmark&\checkmark&&&\checkmark&&&&\checkmark\\
\cline{2-11}
&\cite{m62}(2017) &&\checkmark&&\checkmark&&&&\checkmark&\\
\cline{2-11}
&\cite{m91}(2021) &&\checkmark&&\checkmark&\checkmark&&&\checkmark&\\
\cline{2-11}
&\cite{m92}(2020) &&\checkmark&&\checkmark&\checkmark&&&\checkmark&\\
\cline{2-11}
&\cite{m60}(2019) &&\checkmark&&&&\checkmark&&\checkmark&\\
\hline
\end{tabular}
}
\label{Tabel.Monetary-aware}
\end{table*}

 
\subsection{Physical World-Aware Methods}

All methods discussed so far considered online relationships between users. However, users may have relationships or other things in common through their activities in the physical world. For example, they may live in the same neighbourhood, they may be regulars of the same pub, or may attend events held at the same location. In such cases, users' behaviour may also take into account aspects of their physical world presence. We term such methods as physical world-aware methods and we divide them into two categories: (i) physical relationship-aware methods, where, alongside an online relationship, the physical world relationship between the users is considered to determine the influence of users while messages may propagate through both online and physical world relationships; (ii) location-aware methods, where, the query includes location and the probability that users visit this location is taken into account. In the latter case, the aim is to identify a set of seeds to spread a message so that the number of users visiting the location is maximized. 

\subsubsection{Physical Relationship-Aware Methods}

As mentioned, the main assumption of these methods is that users, in addition to the online world, have some relationships in the physical world too; messages can be propagated in the physical world as well. In order to capture this, the network is modelled as a double-layered graph to show the relationship between users in both the online and physical world. 
Whenever two online users appear to come from the same physical location at the same time, a relationship between these two users in the physical world is implied, hence, an edge in the physical world layer (often termed as offline layer) between these two users is added. Edges in the online layer are processed in the same way as discussed previously. The physical location of users may be determined through online logging information at check-in. 

In~\cite{m74,m73}, if the Euclidean distance between two users over a period of time is less than a defined threshold, these users are considered as neighbours in the physical world layer; clearly the notion of neighbour (and edges) in the physical world layer is dynamic and changes over the time. In~\cite{m74}, to select a seed set, the network is compressed into a single layer and the problem is solved using a simulation-based strategy. The authors also propose a centrality-based method to identify a set of influential seeds covering different parts of the network. In~\cite{m73}, the physical world edges are weighted; according to the ratio of attendance of two users in a common place over a period of time, a weight is assigned to each physical world edge. The two layers are then compressed into a single layer, and the reverse influence sampling method~\cite{RIS1} is used to find influential seeds who can spread a message widely in both the online and physical world layers. 
In~\cite{m95}, the authors take into account the effect of community structure. In their paper, the similarity between a pair of nodes is determined based on the mobility of users in physical world; then, this similarity is used to decompose the network into a set of communities. A simulation-based method and a centrality-based method are proposed to identify the influential nodes in each community. 

In~\cite{m14,m98}, the geographical distance between a pair of users is taken into account to model the behaviour of users. In~\cite{m14}, the authors model the network as a single layered graph; the geographical distance between two users in the physical world along with users' interests are taken into account to model the behaviour of users; the spreading (influence) probability between the users is determined based on their behaviour. Influential nodes are then identified with the help of the reverse influence sampling method~\cite{RIS1}; some heuristics are also proposed to improve the efficiency of the proposed method. 
In~\cite{m98}, the problem is modelled as the identification of a set of influential nodes to maximize the geographical coverage of the locations around the vicinity of the location of the query formulated. The authors propose hierarchical graph modelling alongside some indices to take into account the structural information of the network and location of users. A centrality measure is proposed to identify the geographical coverage of each node and select a set of influential nodes. 

In summary, the model defined in~\cite{m74,m73} 
makes use of a binary value to define whether there is a physical relationship between a pair of nodes. As discussed before, a binary value may not properly capture the behaviour of users, especially the location of users in the physical world which changes dynamically. Thus, in the model defined in~\cite{m14,m98}, a real value is used to determine the physical distance between a pair of users and the spreading probability between them. Yet, neglecting the dynamics of the users' physical location may negatively affect the success of these methods~\cite{m14,m98} in real-world scenarios.

\subsubsection{Location-Aware Methods}

The main assumption in this family of methods is that an event, such as exhibition, commemoration, sale, and so on, is held in a special location. The goal is to spread a publicity message through the network aiming to maximize the number of users visiting the event. 

In~\cite{m65}, according to the location of the event, a target region is included in the query formulated. The goal is to maximize the number of activated users located in the region. In order to find a set of influential nodes, the maximum influence arborescence method~\cite{MIA1} is applied to determine the influence of each node on the nodes in the target region; a set of influential nodes is identified in an iterative manner. Users who are not in the target region but have a short distance to the region are neglected in~\cite{m65}. Thus, instead of determining a region, some studies~\cite{m70,m71,m3,m76}, consider the distance of the users to the location of the event. The goal is to maximize the number of activated users that have a short distance to the location. In~\cite{m70,m71}, for the sake of time-efficiency, a partly offline strategy is adopted. In this strategy a set of sample locations is first determined; the influential set for each sample location is identified and saved offline. When a query comes up, the distance between the sample locations and the event location is considered to determine  influential seeds close to the event location. The problem is further extended in~\cite{m3,m76} by taking additional behavioural information into account. In~\cite{m3}, besides event location, time constraints are also taken into account; the authors apply the reverse influence sampling method~\cite{RIS1} to determine a set of influential seeds. In~\cite{m76}, event location, users' interests and spreading cost are considered; the problem is expressed as a multi-objective optimization problem, which is solved using particle swarm optimization~\cite{PSOalgorithm}.


In~\cite{m72,m75,m68,m19}, the mobility of users is taken into account to determine the probability of attendance of each user in the target region. The goal is to spread the message to the users with high probability of attendance in the region. Users' check-in records are applied in these approaches; the number of times a user checked-in in the region is the measure to estimate the probability of attendance of the user in the region. In~\cite{m72}, the authors discuss how users' check-in behaviour may be applied to model user mobility; they define three Gaussian models to capture mobility. In~\cite{m75}, an extension of the independent cascade model~\cite{tarikh3} is presented to simulate the spreading process based on the event location and the probability of attendance of users in the location. A centrality measure is then defined to determine the influence of each node in this model; the nodes with maximum centrality are chosen as seeds. The role of communities in propagation is taken into account in~\cite{m68}. For this purpose, the maximum influence arborescence method~\cite{MIA1} is applied to determine the influence of each community based on the event location; then, the influential nodes in each community are determined. The influence of communities and nodes are both taken into account to identify a set of influential seeds. In order to speedup the method, the authors also propose an offline tree-based model to determine and ignore the users that have no region intersections with the location defined in the query. In~\cite{m19}, in addition to the users' mobility, users' interests are also taken into account to determine influential nodes; the goal is to activate interested users with a high probability of attendance of the event. A tree-based model is defined to determine the users that have no region intersections with the location defined in the query or the users who are not interested in the event. Then, the maximum influence arborescence method~\cite{MIA1} is extended to determine a set of influential seeds based on  users' interests and probability of attendance of the event. 

In~\cite{m96}, the authors consider a situation where a query aims to spread a marketing message about a sale taking place in specific geographical locations;  each user has a budget to buy some product. Given the budget and geographical distance of each user, the maximum influence arborescence method~\cite{MIA1} is applied to define some rules to select a set of candidate seeds and a centrality measure is proposed to select the influential seeds among them.

Compared to the methods in~\cite{m70,m71,m3,m76}, which determine influential seeds based on the  physical distance of users to the location defined in the query, the methods in~\cite{m72,m75,m68,m19} may capture the behaviour of users more accurately, because they take into account user mobility and the likelihood that users will be in a target location. Nevertheless, the physical location of users may be highly dynamic; tracking and storing all locations for the users over a period of time may bring some processing and storage challenges.

\subsubsection{Summary}

The properties of the physical world-aware methods are summarized in Table~\ref{Tabel.Geographic-aware}. 
The spread of a message and its influence may be affected not only by online interactions but also by physical interactions between users. Therefore, physical relationship-aware methods attempt to model the spreading process and track the influence of users not only on the basis of online interaction but by also incorporating the possible physical interactions between them. Location-aware methods aim to model the IM problem under different scenarios in which users within a specific geographical region are considered to be the target of the spreading process. Overall, although physical world-aware methods attempt to solve the IM problem in a realistic way, taking into account the physical interactions between users and their location imposes an additional complexity to the problem in terms of data modelling and data availability.

\begin{table*}
\centering\footnotesize
\caption{Properties of physical world-aware methods, including applied behavioural features, applied method for influence detection -- Centrality Measure (CM), Spreading Simulation (SS), Reverse Influence Sampling (RIS), and Maximum Influence Arborescence  (MIA) -- and the type of spreading process}
\resizebox{\textwidth}{!}
{
\begin{tabular}{|c|c||c|c|c||c|c|c|c||c|c|}
 \hline
\multirow{2}{4em}{Category} & \multirow{2}{4em}{\centering Paper (year)} & \multicolumn{3}{c||}{Behavioural features} & \multicolumn{4}{c||}{Influence detection}&\multicolumn{2}{c|}{Spreading process}\\ 
\cline{3-11}
&&AM&BV&BE& CM & SS & RIS & MIA&Simple&Complex\\
 \hline
 \multirow{3}{4em}{Physical relationship-aware}& \cite{m74}(2018) &&\checkmark&& \checkmark&\checkmark&&&\checkmark& \\
\cline{2-11}
&\cite{m73}(2018)&&\checkmark&&&&\checkmark&&\checkmark&\checkmark \\
\cline{2-11}
&\cite{m95}(2021) &&\checkmark&&\checkmark&\checkmark&&&\checkmark& \\
\cline{2-11}
& \cite{m14}(2018) &&\checkmark&&&&\checkmark&&\checkmark& \\
\cline{2-11}
& \cite{m98}(2019) &\checkmark&\checkmark&&\checkmark&&&&\checkmark& \\
\hline

\multirow{8}{4em}{Location-aware}& \cite{m65}(2014)&\checkmark&\checkmark&&&&&\checkmark&\checkmark& \\
\cline{2-11}
& \cite{m70}(2016)&\checkmark&\checkmark&&&&\checkmark&\checkmark&\checkmark& \\
\cline{2-11}
& \cite{m71}(2016)&\checkmark&\checkmark&&&&&\checkmark&\checkmark& \\
\cline{2-11}
& \cite{m3}(2016)&\checkmark&\checkmark&&&&\checkmark&&\checkmark&\\
\cline{2-11}
& \cite{m76}(2019) &\checkmark&\checkmark&&&&\checkmark&&\checkmark&\\
\cline{2-11}
& \cite{m75}(2015) &\checkmark&\checkmark&&\checkmark&&&&\checkmark&\\
\cline{2-11}
& \cite{m68}(2018) &\checkmark&\checkmark&&&&&\checkmark&\checkmark&\\
\cline{2-11}
& \cite{m19}(2018)&\checkmark&\checkmark&&&&&\checkmark&\checkmark&\\
\cline{2-11}
& \cite{m96}(2021)&\checkmark&\checkmark&&&&&\checkmark&\checkmark&\\
\hline
\end{tabular}
}
\label{Tabel.Geographic-aware}
\end{table*}


\section{Challenges and Future Directions}
\label{sec_future}

In this paper, we presented a new taxonomy of the methods proposed to solve the IM problem in social networks with a focus on methods that take into account elements of query's features, users' behaviour and relationships' features. 
Research in relation to the IM problem is ongoing and attracts a significant interest as there are various challenges to address. In our view, behaviour-agnostic methods are not always accurate enough to detect influential nodes as they adopt a mechanical approach to link users through a network. Although it has been argued that users' personality may be inferred from online networks \cite{Personalipred1,Personalipred2}, real-world behaviour of individuals is clearly far more complex than what networks may indicate or capture. Behaviour-agnostic methods essentially assume that all users behave the same and their relationships follow a common pattern, an assumption that may not be realistic in real-world scenarios. On the other hand, behaviour-aware methods try to take into account behavioural characteristics to differentiate users and their relationships based on the history of their activities in the network; these characteristics help identify influential users and initiate a successful spreading process more accurately. However, getting hold of and using users' behaviour data may pose complexity and other challenges that will be highlighted next.

\paragraph{Data availability challenges:} Users' behaviour data may not exist for many real-world applications~\cite{challenge1}. Lots of users may be silent or may have no significant activities in a network. Also, they may not be prepared to share important details (e.g., age, gender, location) in their profile, something which may make it difficult to determine the behavioural characteristics of these users and their relationships. As a result, one may deal with a network with incomplete behavioural information in which a considerable number of users and relationships cannot be properly modelled. In addition, users' information may be governed by privacy regulations; making use of such information may not be possible without user consent and full consideration of any privacy issues that may arise. 
  
\paragraph{Data processing challenges:} Obtaining users' behaviour from historical data of users' activities has a number of big data challenges and various preprocessing steps may be necessary before such data is usable~\cite{Challenge2,Challenge3,challenge4}. In general, such data may be noisy, redundant, unstructured or inconsistent. Storing, analysing and processing such data may impose costs and challenges that may be exacerbated in the context of social networks as a result of the increasing and diverse number of user interactions.

\paragraph{Data modelling challenges:} In many behaviour-aware methods, the behaviour of users is modelled using numeric (quantitative) values. However, personality traits may not be simply interpreted using numeric values. For example, determining the opinion of users as a value in $[0,1]$, largely determined by user posts or activities, does not fully capture user attitude. As another example, defining the relationship between a pair of users as a friend or foe relationship is not a straightforward process.

\paragraph{Data dynamicity challenges:} Users and their relationships may have dynamic behavioural characteristics in a network. The structure of the network and the features of the network components may dynamically change over the time. As a result, any data obtained are incidental and may be accompanied by uncertainty. As behavioural features of network components become important, data dynamicity may significantly affect the performance of behaviour-aware methods.

\paragraph{User reaction modelling challenges:} User reaction to suggestions and influence from others is a complicated process which needs to take into account different aspects of personality traits~\cite{SpreadingIssue}. How to model this influence in order to simulate message spreading in a network is a challenging issue. For instance, modelling user reaction to influence from friends or foes may be more complicated than simple binary states currently used by most trust-aware methods.

These challenges point to different directions, which deserve future research.

The first direction that future research will need to consider is the size of the networks. Day by day, the number of users and the volume of data exchanged between the users is increasing. The time complexity of all methods to address the IM problem, particularly when they take into account users' behaviour, needs to be considered. There may be some questions requiring further research in this case: how to propose efficient methods to identify solutions within a factor of the optimal solution, what techniques may be used to compose, effectively and efficiently, offline and online strategies for the IM problem, how to propose and implement highly parallelizable methods that can take advantage of parallel execution.

The second direction is to consider the diversity and dynamicity of the social networks themselves. Different types of networks such as temporal networks, multi-layer networks or dynamic networks are getting popular; there is a need for methods that consider such networks. Some questions to address are: how to take into account the dynamicity of the network structure and user behaviour to find solutions for IM, how to deal with incomplete data and uncertainty in the structural and behavioural information of the networks, how to deal with the diversity of multi-layer networks which are a composition of different social networks.  

The third direction relates to data preparation and how somebody can make use of the data to model the networks efficiently and effectively. This aspect of the problem is not properly addressed in the literature; how to analyse and interpret behavioural data to model a network is a challenge. There are additional questions that may transcend traditional Computer Science boundaries: how to determine the behaviour of users based on their past activities in the network, how to map behavioural data into numeric values to  capture the behaviour of users and relationships.

Finally, information spreading in almost all methods is modelled using a conventional diffusion model. These models may not be the best way to emulate information spreading in reality. Changing the diffusion model may impact the accuracy of the methods significantly. Further research may be needed to assess the impact of users' behaviour in the spreading process and provide suitable diffusion models. A key question is how to take into account the impact that the spreading process may have in users' behaviour, as any such impact may affect subsequent stages of the spreading process. In fact, changes in user behaviour, as they happen, may need to be fed back to the diffusion model itself, something that motivates the need for adaptive and dynamic diffusion models.   

\bibliographystyle{spmpsci}
\bibliography{mybibfile}

\begin{thebibliography}{100}
\providecommand{\url}[1]{{#1}}
\providecommand{\urlprefix}{URL }
\expandafter\ifx\csname urlstyle\endcsname\relax
  \providecommand{\doi}[1]{DOI~\discretionary{}{}{}#1}\else
  \providecommand{\doi}{DOI~\discretionary{}{}{}\begingroup
  \urlstyle{rm}\Url}\fi

\bibitem{m38}
Ahmed, S., Ezeife, C.: Discovering influential nodes from trust network.
\newblock In: Proceedings of the 28th Annual {ACM} Symposium on Applied
  Computing, pp. 121--128. ACM (2013)

\bibitem{Surv10}
Al-Garadi, M.A., Varathan, K.D., Ravana, S.D., Ahmed, E., Mujtaba, G., Khan,
  M.U.S., Khan, S.U.: Analysis of online social network connections for
  identification of influential users: Survey and open research issues.
\newblock ACM Computing Surveys (CSUR) \textbf{51}(1), 16 (2018)

\bibitem{SpreadingIssue}
Aral, S., Walker, D.: Identifying influential and susceptible members of social
  networks.
\newblock Science \textbf{337}(6092), 337--341 (2012)

\bibitem{Surv9}
Arora, A., Galhotra, S., Ranu, S.: Debunking the myths of influence
  maximization: An in-depth benchmarking study.
\newblock In: Proceedings of the 2017 ACM International Conference on
  Management of Data, pp. 651--666. ACM (2017)

\bibitem{m17}
Aslay, C., Barbieri, N., Bonchi, F., Baeza-Yates, R.A.: Online topic-aware
  influence maximization queries.
\newblock In: Proceedings of the 17th International Conference on Extending
  Database Technology (EDBT), pp. 295--306 (2014)

\bibitem{m54}
Banerjee, S., Jenamani, M., Pratihar, D.K.: Combim: A community-based solution
  approach for the budgeted influence maximization problem.
\newblock Expert Systems with Applications \textbf{125}, 1--13 (2019)

\bibitem{m92}
Banerjee, S., Jenamani, M., Pratihar, D.K.: Maximizing the earned benefit in an
  incentivized social networking environment: a community-based approach.
\newblock Journal of Ambient Intelligence and Humanized Computing
  \textbf{11}(6), 2539--2555 (2020)

\bibitem{surv13}
Banerjee, S., Jenamani, M., Pratihar, D.K.: A survey on influence maximization
  in a social network.
\newblock Knowledge and Information Systems \textbf{62}(9), 3417--3455 (2020)

\bibitem{m91}
Banerjee, S., Jenamani, M., Pratihar, D.K.: Earned benefit maximization in
  social networks under budget constraint.
\newblock Expert Systems with Applications \textbf{169}, 114346 (2021)

\bibitem{m93}
Banerjee, S., Pal, B., Jenamani, M.: Budgeted influence maximization with tags
  in social networks.
\newblock In: International Conference on Web Information Systems Engineering,
  pp. 141--152. Springer (2020)

\bibitem{m10}
Barbieri, N., Bonchi, F., Manco, G.: Topic-aware social influence propagation
  models.
\newblock Knowledge and Information Systems \textbf{37}(3), 555--584 (2013)

\bibitem{Challenge2}
Bello-Orgaz, G., Jung, J.J., Camacho, D.: Social big data: Recent achievements
  and new challenges.
\newblock Information Fusion \textbf{28}, 45--59 (2016)

\bibitem{Surv11}
Bian, R., Koh, Y.S., Dobbie, G., Divoli, A.: Identifying top-k nodes in social
  networks: A survey.
\newblock ACM Computing Surveys (CSUR) \textbf{52}(1), 22 (2019)

\bibitem{RIS3}
Borgs, C., Brautbar, M., Chayes, J., Lucier, B.: Maximizing social influence in
  nearly optimal time.
\newblock In: Proceedings of the 25th Annual ACM-SIAM Symposium on Discrete
  Algorithms, pp. 946--957. SIAM (2014)

\bibitem{RIS1}
Borgs, C., Brautbar, M., Chayes, J.T., Lucier, B.: Influence maximization in
  social networks: Towards an optimal algorithmic solution.
\newblock arXiv preprint arXiv:1212.0884  (2012)

\bibitem{TModel1}
Borodin, A., Filmus, Y., Oren, J.: Threshold models for competitive influence
  in social networks.
\newblock In: International workshop on internet and network economics, pp.
  539--550. Springer (2010)

\bibitem{PageRank}
Brin, S., Page, L.: Reprint of: The anatomy of a large-scale hypertextual web
  search engine.
\newblock Computer Networks \textbf{56}(18), 3825--3833 (2012)

\bibitem{m18}
Calio, A., Interdonato, R., Pulice, C., Tagarelli, A.: Topology-driven
  diversity for targeted influence maximization with application to user
  engagement in social networks.
\newblock IEEE Transactions on Knowledge and Data Engineering \textbf{30}(12),
  2421--2434 (2018)

\bibitem{m79}
Caliò, A., Tagarelli, A.: Attribute based diversification of seeds for
  targeted influence maximization.
\newblock Information Sciences \textbf{546}, 1273--1305 (2021)

\bibitem{Cmodel1}
Carnes, T., Nagarajan, C., Wild, S.M., Van~Zuylen, A.: Maximizing influence in
  a competitive social network: a follower's perspective.
\newblock In: Proceedings of the 9th International Conference on Electronic
  Commerce, pp. 351--360. ACM (2007)

\bibitem{m15}
Chen, S., Fan, J., Li, G., Feng, J., Tan, K.l., Tang, J.: Online topic-aware
  influence maximization.
\newblock Proceedings of the {VLDB} Endowment \textbf{8}(6), 666--677 (2015)

\bibitem{m34}
Chen, S., He, K.: Influence maximization on signed social networks with
  integrated pagerank.
\newblock In: 2015 IEEE International Conference on Smart
  City/SocialCom/SustainCom, pp. 289--292. IEEE (2015)

\bibitem{m39}
Chen, W., Collins, A., Cummings, R., Ke, T., Liu, Z., Rincon, D., Sun, X.,
  Wang, Y., Wei, W., Yuan, Y.: Influence maximization in social networks when
  negative opinions may emerge and propagate.
\newblock In: Proceedings of the 2011 {SIAM} International Conference on Data
  Mining, pp. 379--390. SIAM (2011)

\bibitem{Surv8}
Chen, W., Lakshmanan, L.V., Castillo, C.: Information and influence propagation
  in social networks.
\newblock Synthesis Lectures on Data Management \textbf{5}(4), 1--177 (2013)

\bibitem{m9}
Chen, W., Lin, T., Yang, C.: Real-time topic-aware influence maximization using
  preprocessing.
\newblock Computational Social Networks \textbf{3}(1), 8 (2016)

\bibitem{MIA1}
Chen, W., Wang, C., Wang, Y.: Scalable influence maximization for prevalent
  viral marketing in large-scale social networks.
\newblock In: Proceedings of the 16th ACM SIGKDD International Conference on
  Knowledge Discovery and Data Mining, pp. 1029--1038. ACM (2010)

\bibitem{SIMB1}
Chen, W., Wang, Y., Yang, S.: Efficient influence maximization in social
  networks.
\newblock In: Proceedings of the 15th ACM SIGKDD International Conference on
  Knowledge Discovery and Data Mining, pp. 199--208. ACM (2009)

\bibitem{m95}
Chen, X., Deng, L., Zhao, Y., Zhou, X., Zheng, K.: Community-based influence
  maximization in location-based social network.
\newblock World Wide Web \textbf{24}(6), 1903--1928 (2021)

\bibitem{m59}
Chiesse, R., Figueiredo, D.R., Antonio, A.d.A., Ziviani, A., Niter{\'o}i, R.,
  Petr{\'o}polis, R.: Evaluation of epidemic seeding strategies under variable
  node costs.
\newblock In: WPerformance 2014 Workshop, XXXIV Congresso Nacional da Sociedade
  Brasileira de Computa\c{c}\~ao (CSBC), Brazil (2014)

\bibitem{challenge4}
Cuomo, S., Maiorano, F.: Social network data analysis and mining applications
  for the internet of data.
\newblock Concurrency and Computation: Practice and Experience \textbf{30}(15),
  e4527 (2018)

\bibitem{Surv4}
Das, K., Samanta, S., Pal, M.: Study on centrality measures in social networks:
  a survey.
\newblock Social Network Analysis and Mining \textbf{8}(1), 13 (2018)

\bibitem{Prediction2}
De~Salve, A., Mori, P., Guidi, B., Ricci, L., Pietro, R.D.: Predicting
  influential users in online social network groups.
\newblock ACM Transactions on Knowledge Discovery from Data \textbf{15}(3)
  (2021)

\bibitem{tarikh1}
Domingos, P., Richardson, M.: Mining the network value of customers.
\newblock In: Proceedings of the 7th ACM SIGKDD International Conference on
  Knowledge Discovery and Data Mining, pp. 57--66. ACM (2001)

\bibitem{m56}
Du, N., Liang, Y., Balcan, M.F., Song, L.: Budgeted influence maximization for
  multiple products.
\newblock arXiv preprint arXiv:1312.2164  (2013)

\bibitem{PSOalgorithm}
Eberhart, R., Kennedy, J.: A new optimizer using particle swarm theory.
\newblock In: MHS'95. Proceedings of the 6th International Symposium on Micro
  Machine and Human Science, pp. 39--43. Ieee (1995)

\bibitem{Betweenness}
Freeman, L.C.: A set of measures of centrality based on betweenness.
\newblock Sociometry \textbf{40}(1), 35--41 (1977)

\bibitem{Degree}
Freeman, L.C.: Centrality in social networks conceptual clarification.
\newblock Social Networks \textbf{1}(3), 215--239 (1978)

\bibitem{m48}
Gionis, A., Terzi, E., Tsaparas, P.: Opinion maximization in social networks.
\newblock In: Proceedings of the 2013 SIAM International Conference on Data
  Mining, pp. 387--395. SIAM (2013)

\bibitem{Personalipred1}
Golbeck, J., Robles, C., Edmondson, M., Turner, K.: Predicting personality from
  twitter.
\newblock In: 2011 IEEE 3rd international conference on privacy, security, risk
  and trust and 2011 IEEE third international conference on social computing,
  pp. 149--156. IEEE (2011)

\bibitem{Personalipred2}
Golbeck, J., Robles, C., Turner, K.: Predicting personality with social media.
\newblock In: CHI'11 extended abstracts on human factors in computing systems,
  pp. 253--262. ACM (2011)

\bibitem{Cmodel2}
Goldenberg, J., Libai, B., Muller, E.: Using complex systems analysis to
  advance marketing theory development: Modeling heterogeneity effects on new
  product growth through stochastic cellular automata.
\newblock Academy of Marketing Science Review \textbf{9}(3), 1--18 (2001)

\bibitem{Prediction1}
Gong, Q., Chen, Y., He, X., Xiao, Y., Hui, P., Wang, X., Fu, X.: Cross-site
  prediction on social influence for cold-start users in online social
  networks.
\newblock ACM Transactions on the Web \textbf{15}(2) (2021)

\bibitem{Tmodel2}
Granovetter, M.: Threshold models of collective behavior.
\newblock American Journal of Sociology \textbf{83}(6), 1420--1443 (1978)

\bibitem{m96}
Gu, Y., Yao, X., Liang, G., Gu, C., Huang, H.: Efficient budget-distance-aware
  influence maximization in geo-social network.
\newblock In: International Conference on Wireless Algorithms, Systems, and
  Applications, pp. 282--290. Springer (2021)

\bibitem{Surv6}
Guille, A., Hacid, H., Favre, C., Zighed, D.A.: Information diffusion in online
  social networks: A survey.
\newblock ACM Sigmod Record \textbf{42}(2), 17--28 (2013)

\bibitem{m53}
G{\"u}ney, E.: On the optimal solution of budgeted influence maximization
  problem in social networks.
\newblock Operational Research \textbf{19}(3), 817--831 (2019)

\bibitem{Surv14}
Hafiene, N., Karoui, W., {Ben Romdhane}, L.: Influential nodes detection in
  dynamic social networks: A survey.
\newblock Expert Systems with Applications \textbf{159}, 113642 (2020)

\bibitem{m94}
Han, K., Wu, B., Tang, J., Cui, S., Aslay, C., Lakshmanan, L.V.: Efficient and
  Effective Algorithms for Revenue Maximization in Social Advertising, p.
  671–684.
\newblock Association for Computing Machinery, New York, NY, USA (2021)

\bibitem{m50}
Han, S., Zhuang, F., He, Q., Shi, Z.: Balanced seed selection for budgeted
  influence maximization in social networks.
\newblock In: Pacific-Asia Conference on Knowledge Discovery and Data Mining,
  pp. 65--77. Springer (2014)

\bibitem{m31}
He, J.S., Kaur, H., Talluri, M.: Positive opinion influential node set
  selection for social networks: Considering both positive and negative
  relationships.
\newblock In: Wireless Communications, Networking and Applications, pp.
  935--948. Springer (2016)

\bibitem{m86}
He, Q., Fang, H., Zhang, J., Wang, X.: Dynamic opinion maximization in social
  networks.
\newblock IEEE Transactions on Knowledge and Data Engineering \textbf{35}(1),
  350--361 (2023)

\bibitem{m90}
He, Q., Lv, Y., Wang, X., Li, J., Huang, M., Ma, L., Cai, Y.: Reinforcement
  learning based dynamic opinion maximization framework in signed social
  networks.
\newblock IEEE Transactions on Cognitive and Developmental Systems pp. 1--11
  (2022)

\bibitem{m89}
He, Q., Sun, L., Wang, X., Wang, Z., Huang, M., Yi, B., Wang, Y., Ma, L.:
  Positive opinion maximization in signed social networks.
\newblock Information Sciences \textbf{558}, 34--49 (2021)

\bibitem{m85}
He, Q., Wang, X., Huang, M., Yi, B.: Multi-stage opinion maximization in social
  networks.
\newblock Neural Computing and Applications \textbf{33}(19), 12367--12380
  (2021)

\bibitem{m44}
Hosseini-Pozveh, M., Zamanifar, K., Naghsh-Nilchi, A.R., Dolog, P.: Maximizing
  the spread of positive influence in signed social networks.
\newblock Intelligent Data Analysis \textbf{20}(1), 199--218 (2016)

\bibitem{m98}
Hosseinpour, M., Malek, M.R., Claramunt, C.: Socio-spatial influence
  maximization in location-based social networks.
\newblock Future Generation Computer Systems \textbf{101}, 304--314 (2019)

\bibitem{challenge1}
Jalili, M., Perc, M.: Information cascades in complex networks.
\newblock Journal of Complex Networks \textbf{5}(5), 665--693 (2017)

\bibitem{Surv15}
Jaouadi, M., Ben~Romdhane, L.: Influence maximization problem in social
  networks: An overview.
\newblock In: 2019 IEEE/ACS 16th International Conference on Computer Systems
  and Applications (AICCSA), pp. 1--8 (2019)

\bibitem{m87}
Ju, W., Chen, L., Li, B., Liu, W., Sheng, J., Wang, Y.: A new algorithm for
  positive influence maximization in signed networks.
\newblock Information Sciences \textbf{512}, 1571--1591 (2020)

\bibitem{m20}
Ke, X., Khan, A., Cong, G.: Finding seeds and relevant tags jointly: For
  targeted influence maximization in social networks.
\newblock In: Proceedings of the 2018 International Conference on Management of
  Data, p. 1097–1111. Association for Computing Machinery (2018)

\bibitem{tarikh3}
Kempe, D., Kleinberg, J., Tardos, {\'E}.: Maximizing the spread of influence
  through a social network.
\newblock In: Proceedings of the 9th ACM SIGKDD International Conference on
  Knowledge Discovery and Data Mining, pp. 137--146. ACM (2003)

\bibitem{K-shell}
Kitsak, M., Gallos, L.K., Havlin, S., Liljeros, F., Muchnik, L., Stanley, H.E.,
  Makse, H.A.: Identification of influential spreaders in complex networks.
\newblock Nature Physics \textbf{6}(11), 888 (2010)

\bibitem{m30}
Lei, W., Yang, Q., Wang, H.: Positive influence maximization algorithm based on
  three degrees of influence.
\newblock In: International Conference on Intelligent Data Engineering and
  Automated Learning, pp. 503--514. Springer (2016)

\bibitem{m25}
Li, D., Wang, C., Zhang, S., Zhou, G., Chu, D., Wu, C.: Positive influence
  maximization in signed social networks based on simulated annealing.
\newblock Neurocomputing \textbf{260}, 69--78 (2017)

\bibitem{m36}
Li, D., Xu, Z.M., Chakraborty, N., Gupta, A., Sycara, K., Li, S.: Polarity
  related influence maximization in signed social networks.
\newblock PloS one \textbf{9}(7), e102199 (2014)

\bibitem{m1}
Li, F.H., Li, C.T., Shan, M.K.: Labeled influence maximization in social
  networks for target marketing.
\newblock In: 2011 IEEE International Conference on Privacy, Security, Risk and
  Trust and IEEE International Conference on Social Computing, pp. 560--563.
  IEEE (2011)

\bibitem{m65}
Li, G., Chen, S., Feng, J., Tan, K.l., Li, W.s.: Efficient location-aware
  influence maximization.
\newblock In: Proceedings of the 2014 ACM SIGMOD International Conference on
  Management of Data, pp. 87--98. ACM (2014)

\bibitem{m14}
Li, J., Cai, T., Mian, A., Li, R.H., Sellis, T., Yu, J.X.: Holistic influence
  maximization for targeted advertisements in spatial social networks.
\newblock In: 2018 IEEE 34th International Conference on Data Engineering
  (ICDE), pp. 1340--1343. IEEE (2018)

\bibitem{m84}
Li, L., Liu, Y., Zhou, Q., Yang, W., Yuan, J.: Targeted influence maximization
  under a multifactor-based information propagation model.
\newblock Information Sciences \textbf{519}, 124--140 (2020)

\bibitem{IM1}
Li, M., Wang, Z., Han, Q.L., Taylor, S.J.E., Li, K., Liao, X., Liu, X.:
  Influence maximization in multiagent systems by a graph embedding method:
  Dealing with probabilistically unstable links.
\newblock IEEE Transactions on Cybernetics pp. 1--13 (2022)

\bibitem{m73}
Li, S., Han, K., Zhang, J.: Exploring influence maximization in location-based
  social networks.
\newblock In: International Conference on Collaborative Computing: Networking,
  Applications and Worksharing, pp. 92--111. Springer (2018)

\bibitem{m68}
Li, X., Cheng, X., Su, S., Sun, C.: Community-based seeds selection algorithm
  for location aware influence maximization.
\newblock Neurocomputing \textbf{275}, 1601--1613 (2018)

\bibitem{m46}
Li, Y., Chen, W., Wang, Y., Zhang, Z.L.: Influence diffusion dynamics and
  influence maximization in social networks with friend and foe relationships.
\newblock In: Proceedings of the 6th ACM International Conference on Web Search
  and Data Mining, pp. 657--666. ACM (2013)

\bibitem{Surv2}
Li, Y., Fan, J., Wang, Y., Tan, K.L.: Influence maximization on social graphs:
  A survey.
\newblock IEEE Transactions on Knowledge and Data Engineering \textbf{30}(10),
  1852--1872 (2018)

\bibitem{m82}
Li, Y., Gan, X., Fu, L., Tian, X., Qin, Z., Zhou, Y.: Conformity-aware
  influence maximization with user profiles.
\newblock In: 2018 10th International Conference on Wireless Communications and
  Signal Processing (WCSP), pp. 1--6 (2018)

\bibitem{m2}
Li, Y., Zhang, D., Tan, K.L.: Real-time targeted influence maximization for
  online advertisements.
\newblock Proceedings of the VLDB Endowment \textbf{8}(10), 1070--1081 (2015)

\bibitem{m29}
Liang, W., Shen, C., Li, X., Nishide, R., Piumarta, I., Takada, H.: Influence
  maximization in signed social networks with opinion formation.
\newblock IEEE Access  (2019)

\bibitem{m88}
Liang, W., Shen, C., Li, X., Nishide, R., Piumarta, I., Takada, H.: Influence
  maximization in signed social networks with opinion formation.
\newblock IEEE Access \textbf{7}, 68837--68852 (2019)

\bibitem{m23-1}
Liu, S., Jiang, C., Lin, Z., Ding, Y., Duan, R., Xu, Z.: Identifying effective
  influencers based on trust for electronic word-of-mouth marketing: A
  domain-aware approach.
\newblock Information Sciences \textbf{306}, 34--52 (2015)

\bibitem{m81}
Lohia, P., Kannan, K., Rai, K., Bedathur, S.: Ranking marginal influencers in a
  target-labeled network.
\newblock In: Proceedings of the 7th ACM IKDD CoDS and 25th COMAD, p.
  254–260. Association for Computing Machinery (2020)

\bibitem{Surv1}
L{\"u}, L., Chen, D., Ren, X.L., Zhang, Q.M., Zhang, Y.C., Zhou, T.: Vital
  nodes identification in complex networks.
\newblock Physics Reports \textbf{650}, 1--63 (2016)

\bibitem{H-index}
L{\"u}, L., Zhou, T., Zhang, Q.M., Stanley, H.E.: The h-index of a network node
  and its relation to degree and coreness.
\newblock Nature Communications \textbf{7}, 10168 (2016)

\bibitem{m61}
Lu, W., Lakshmanan, L.V.: Profit maximization over social networks.
\newblock In: 2012 IEEE 12th International Conference on Data Mining, pp.
  479--488. IEEE (2012)

\bibitem{Challenge3}
Manovich, L.: Trending: The promises and the challenges of big social data.
\newblock Debates in the digital humanities \textbf{2}, 460--475 (2011)

\bibitem{m4}
Mochalova, A., Nanopoulos, A.: A targeted approach to viral marketing.
\newblock Electronic Commerce Research and Applications \textbf{13}(4),
  283--294 (2014)

\bibitem{m37}
Mohamadi-Baghmolaei, R., Mozafari, N., Hamzeh, A.: Trust based latency aware
  influence maximization in social networks.
\newblock Engineering Applications of Artificial Intelligence \textbf{41},
  195--206 (2015)

\bibitem{m41}
Nazemian, A., Taghiyareh, F.: Influence maximization in independent cascade
  model with positive and negative word of mouth.
\newblock In: 6th International Symposium on Telecommunications (IST), pp.
  854--860. IEEE (2012)

\bibitem{m51}
Nguyen, H., Zheng, R.: On budgeted influence maximization in social networks.
\newblock IEEE Journal on Selected Areas in Communications \textbf{31}(6),
  1084--1094 (2013)

\bibitem{m16}
Nguyen, H.T., Dinh, T.N., Thai, M.T.: Cost-aware targeted viral marketing in
  billion-scale networks.
\newblock In: IEEE INFOCOM 2016-The 35th Annual IEEE International Conference
  on Computer Communications, pp. 1--9. IEEE (2016)

\bibitem{m80}
Padmanabhan, M.R., Somisetty, N., Basu, S., Pavan, A.: Influence maximization
  in social networks with non-target constraints.
\newblock In: 2018 IEEE International Conference on Big Data, pp. 771--780
  (2018)

\bibitem{Surv3}
Peng, S., Zhou, Y., Cao, L., Yu, S., Niu, J., Jia, W.: Influence analysis in
  social networks: a survey.
\newblock Journal of Network and Computer Applications \textbf{106}, 17--32
  (2018)

\bibitem{tarikh2}
Richardson, M., Domingos, P.: Mining knowledge-sharing sites for viral
  marketing.
\newblock In: Proceedings of the 8th ACM SIGKDD International Conference on
  Knowledge Discovery and Data Mining, pp. 61--70. ACM (2002)

\bibitem{Closeness}
Sabidussi, G.: The centrality index of a graph.
\newblock Psychometrika \textbf{31}(4), 581--603 (1966)

\bibitem{m33}
Shafaei, M., Jalili, M.: Community structure and information cascade in signed
  networks.
\newblock New Generation Computing \textbf{32}(3-4), 257--269 (2014)

\bibitem{m7}
Singh, S.S., Kumar, A., Singh, K., Biswas, B.: C2im: Community based
  context-aware influence maximization in social networks.
\newblock Physica A: Statistical Mechanics and its Applications \textbf{514},
  796--818 (2019)

\bibitem{Surv7}
Singh, S.S., Singh, K., Kumar, A., Shakya, H.K., Biswas, B.: A survey on
  information diffusion models in social networks.
\newblock In: International Conference on Advanced Informatics for Computing
  Research, pp. 426--439. Springer (2018)

\bibitem{m3}
Song, C., Hsu, W., Lee, M.L.: Targeted influence maximization in social
  networks.
\newblock In: Proceedings of the 25th ACM International Conference on
  Information and Knowledge Management, pp. 1683--1692. ACM (2016)

\bibitem{m58}
de~Souza, R., Figueiredo, D., Rocha, A.d.A., Ziviani, A.: Efficient network
  seeding under variable node cost and limited budget for social networks.
\newblock Information Sciences \textbf{514}, 369--384 (2020)

\bibitem{m12}
Srinivasan, B.V., Anandhavelu, N., Dalal, A., Yenugula, M., Srikanthan, P.,
  Layek, A.: Topic-based targeted influence maximization.
\newblock In: 2014 6th International Conference on Communication Systems and
  Networks (COMSNETS), pp. 1--6. IEEE (2014)

\bibitem{m42}
Srivastava, A., Chelmis, C., Prasanna, V.K.: Social influence computation and
  maximization in signed networks with competing cascades.
\newblock In: 2015 IEEE/ACM International Conference on Advances in Social
  Networks Analysis and Mining (ASONAM), pp. 41--48. IEEE (2015)

\bibitem{m19}
Su, S., Li, X., Cheng, X., Sun, C.: Location-aware targeted influence
  maximization in social networks.
\newblock Journal of the Association for Information Science and Technology
  \textbf{69}(2), 229--241 (2018)

\bibitem{m62}
Tang, J., Tang, X., Yuan, J.: Profit maximization for viral marketing in online
  social networks: Algorithms and analysis.
\newblock IEEE Transactions on Knowledge and Data Engineering \textbf{30}(6),
  1095--1108 (2017)

\bibitem{RIS2}
Tang, Y., Xiao, X., Shi, Y.: Influence maximization: Near-optimal time
  complexity meets practical efficiency.
\newblock In: Proceedings of the 2014 ACM SIGMOD International Conference on
  Management of Data, pp. 75--86. ACM (2014)

\bibitem{Surv5}
Tejaswi, V., Bindu, P., Thilagam, P.S.: Diffusion models and approaches for
  influence maximization in social networks.
\newblock In: 2016 International Conference on Advances in Computing,
  Communications and Informatics (ICACCI), pp. 1345--1351. IEEE (2016)

\bibitem{m13}
Tejaswi, V., Bindu, P., Thilagam, P.S.: Target specific influence maximization:
  An approach to maximize adoption in labeled social networks.
\newblock In: 2017 9th International Conference on Communication Systems and
  Networks (COMSNETS), pp. 542--547. IEEE (2017)

\bibitem{m83}
Tian, S., Mo, S., Wang, L., Peng, Z.: Deep reinforcement learning-based
  approach to tackle topic-aware influence maximization.
\newblock Data Science and Engineering \textbf{5}(1), 1--11 (2020)

\bibitem{m45}
Wang, F., Jiang, W., Li, X., Wang, G.: Maximizing positive influence spread in
  online social networks via fluid dynamics.
\newblock Future Generation Computer Systems \textbf{86}, 1491--1502 (2018)

\bibitem{m40}
Wang, F., Wang, G., Xie, D.: Maximizing the spread of positive influence under
  lt-mla model.
\newblock In: Asia-Pacific Services Computing Conference, pp. 450--463.
  Springer (2016)

\bibitem{m28}
Wang, H., Yang, Q., Fang, L., Lei, W.: Maximizing positive influence in signed
  social networks.
\newblock In: International Conference on Cloud Computing and Security, pp.
  356--367. Springer (2015)

\bibitem{m76}
Wang, L., Yu, Z., Xiong, F., Yang, D., Pan, S., Yan, Z.: Influence spread in
  geo-social networks: A multiobjective optimization perspective.
\newblock IEEE Transactions on Cybernetics  (2019)

\bibitem{m71}
Wang, X., Zhang, Y., Zhang, W., Lin, X.: Distance-aware influence maximization
  in geo-social network.
\newblock In: ICDE, pp. 1--12 (2016)

\bibitem{m70}
Wang, X., Zhang, Y., Zhang, W., Lin, X.: Efficient distance-aware influence
  maximization in geo-social networks.
\newblock IEEE Transactions on Knowledge and Data Engineering \textbf{29}(3),
  599--612 (2016)

\bibitem{m63}
Wang, Y., Vasilakos, A.V., Jin, Q., Ma, J.: Pprank: economically selecting
  initial users for influence maximization in social networks.
\newblock IEEE Systems Journal \textbf{11}(4), 2279--2290 (2015)

\bibitem{m11}
Wen, Y.T., Peng, W.C., Shuai, H.H.: Maximizing social influence on target
  users.
\newblock In: Pacific-Asia Conference on Knowledge Discovery and Data Mining,
  pp. 701--712. Springer (2018)

\bibitem{m5}
White, S., Smyth, P.: Algorithms for estimating relative importance in
  networks.
\newblock In: Proceedings of the 9th ACM SIGKDD International Conference on
  Knowledge Discovery and Data Mining, pp. 266--275. ACM (2003)

\bibitem{Surv12}
Yang, Y., Pei, J.: Influence analysis in evolving networks: A survey.
\newblock IEEE Transactions on Knowledge and Data Engineering  (2019)

\bibitem{m74}
Yang, Y., Xu, Y., Wang, E., Lou, K., Luan, D.: Exploring influence maximization
  in online and offline double-layer propagation scheme.
\newblock Information Sciences \textbf{450}, 182--199 (2018)

\bibitem{m57}
Yu, Q., Li, H., Liao, Y., Cui, S.: Fast budgeted influence maximization over
  multi-action event logs.
\newblock IEEE Access \textbf{6}, 14367--14378 (2018)

\bibitem{ehc}
Zareie, A., Sheikhahmadi, A.: Ehc: Extended h-index centrality measure for
  identification of users’ spreading influence in complex networks.
\newblock Physica A: Statistical Mechanics and its Applications \textbf{514},
  141--155 (2019)

\bibitem{ERM}
Zareie, A., Sheikhahmadi, A., Fatemi, A.: Influential nodes ranking in complex
  networks: An entropy-based approach.
\newblock Chaos, Solitons \& Fractals \textbf{104}, 485--494 (2017)

\bibitem{IMUD}
Zareie, A., Sheikhahmadi, A., Jalili, M.: Identification of influential users
  in social networks based on users’ interest.
\newblock Information Sciences \textbf{493}, 217--231 (2019)

\bibitem{m43}
Zhang, H., Dinh, T.N., Thai, M.T.: Maximizing the spread of positive influence
  in online social networks.
\newblock In: 2013 IEEE 33rd International Conference on Distributed Computing
  Systems, pp. 317--326. IEEE (2013)

\bibitem{m64}
Zhang, H., Zhang, H., Kuhnle, A., Thai, M.T.: Profit maximization for multiple
  products in online social networks.
\newblock In: IEEE INFOCOM 2016-The 35th Annual IEEE International Conference
  on Computer Communications, pp. 1--9. IEEE (2016)

\bibitem{IM2}
Zhang, Y., Guo, J., Yang, W., Wu, W.: Supplementary influence maximization
  problem in social networks.
\newblock IEEE Transactions on Computational Social Systems pp. 1--11 (2023)

\bibitem{m6}
Zhou, J., Zhang, Y., Cheng, J.: Preference-based mining of top-k influential
  nodes in social networks.
\newblock Future Generation Computer Systems \textbf{31}, 40--47 (2014)

\bibitem{m75}
Zhou, T., Cao, J., Liu, B., Xu, S., Zhu, Z., Luo, J.: Location-based influence
  maximization in social networks.
\newblock In: Proceedings of the 24th ACM International Conference on
  Information and Knowledge Management, pp. 1211--1220. ACM (2015)

\bibitem{m60}
Zhu, J., Ghosh, S., Wu, W., Gao, C.: Profit maximization under group influence
  model in social networks.
\newblock In: International Conference on Computational Data and Social
  Networks, pp. 108--119. Springer (2019)

\bibitem{m72}
Zhu, W.Y., Peng, W.C., Chen, L.J., Zheng, K., Zhou, X.: Modeling user mobility
  for location promotion in location-based social networks.
\newblock In: Proceedings of the 21st ACM SIGKDD International Conference on
  Knowledge Discovery and Data Mining, pp. 1573--1582. ACM (2015)

\end{thebibliography}
\end{document}